\documentclass[11pt]{article}
\usepackage{bm}
\usepackage{geometry}
\usepackage{fullpage}
\usepackage{url}
\usepackage[english]{babel}
\usepackage{graphicx}
\usepackage{color}
\usepackage{rotating}
\usepackage{tabularx}
\usepackage{natbib}
\usepackage{amsmath}
\usepackage{comment}
\includecomment{comment}
\usepackage{multirow}
\usepackage{hhline}
\usepackage[dvipsnames]{xcolor}
\usepackage[unicode=true,pdfusetitle,
 bookmarks=true,bookmarksnumbered=true,bookmarksopen=true,bookmarksopenlevel=2,
 breaklinks=true,pdfborder={0 0 1},backref=false,colorlinks=true,citecolor=blue,urlcolor=blue,linkcolor=blue]
 {hyperref}

\begin{document}








\begin{center}
{\bf A note on the modeling of the effects of experimental time \\ in psycholinguistic experiments}
\end{center}

\begin{center}

R. Harald Baayen$^1$, Matteo Fasiolo$^2$, Simon Wood$^3$, and Yu-Ying Chuang$^1$, \\
1: Quantitative Linguistics, Department of Linguistics, University of T\"{u}bingen, 2: Statistical Science, School of Mathematics, University of Bristol, 3: Computational Statistics, School of Mathematics, University of Edinburgh \footnote{This research was supported by the European Research Council (WIDE-742545 to the first author), and by the Center for Interdisciplinary Research, Bielefeld (ZiF)/Cooperation Group ``Statistical models for psychological and linguistic data''.  The authors are indebted to Douglas Bates for comments on previous versions of this paper. 
}
\end{center}
\ \\

\begin{center}
  {\bf Abstract} \\ 
\end{center}

\noindent
\citet{BarrGam:2020} called attention to problems that arise when chronometric experiments implementing specific factorial designs are analysed with the generalized additive mixed model (GAMM), using factor smooths to capture trial-to-trial dependencies. {\color{black} From a series of simulations incorporating such dependencies, they conclude that GAMMs are inappropriate for between-subject designs. They argue that in addition GAMMs come with too many modeling possibilities (often referred to as `researcher degrees of freedom'), and advise using the linear mixed model (LMM) instead. As clarified by the title of their paper, their conclusion is: ``Using gamms to model trial-by-trial fluctuations in experimental data:  More risks but hardly any benefit''.}  

We address the questions raised by \citet{BarrGam:2020}, who clearly demonstrated that problems can indeed arise when using factor smooths in combination with factorial designs. We show that the problem does not arise when using by-smooths.  Furthermore, we have traced a bug in the implementation of factor smooths in the {\bf mgcv} package, which will have been removed from version 1.8-36 onwards. 

To illustrate that GAMMs now produce correct estimates, we report simulation studies implementing different by-subject longitudinal effects.  The maximal LMM emerges as slightly conservative compared to GAMMs, and GAMMs provide estimated coefficients that {\color{black} can be} less variable across simulation runs.  We also discuss two datasets where time-varying effects interact with {\color{black} numerical predictors in a theoretically informative way.}  

{\color{black} Furthermore, we argue that the  wide range of tools that GAMMs make available to researcher across all domains of scientific inquiry do not come with uncontrolled researcher degrees of freedom once confronted with a specific psycholinguistic datasets. We also introduce a distinction between replicable and non-replicable non-linear effects.}

We conclude that GAMMs are an excellent and reliable tool for understanding {\color{black} experimental data, including } chronometric data with time-varying effects.  
\\ \ \\

\begin{flushleft}
{\bf Keywords:} 
\end{flushleft}
generalized additive mixed model, factor smooths, by smooths, factorial designs, \\ time-varying effects, random effects, researcher degrees of freedom, replication \\ \ \\

\newpage
\section{Introduction}

\begin{figure}[b]
\centering
\includegraphics[width=0.8\textwidth]{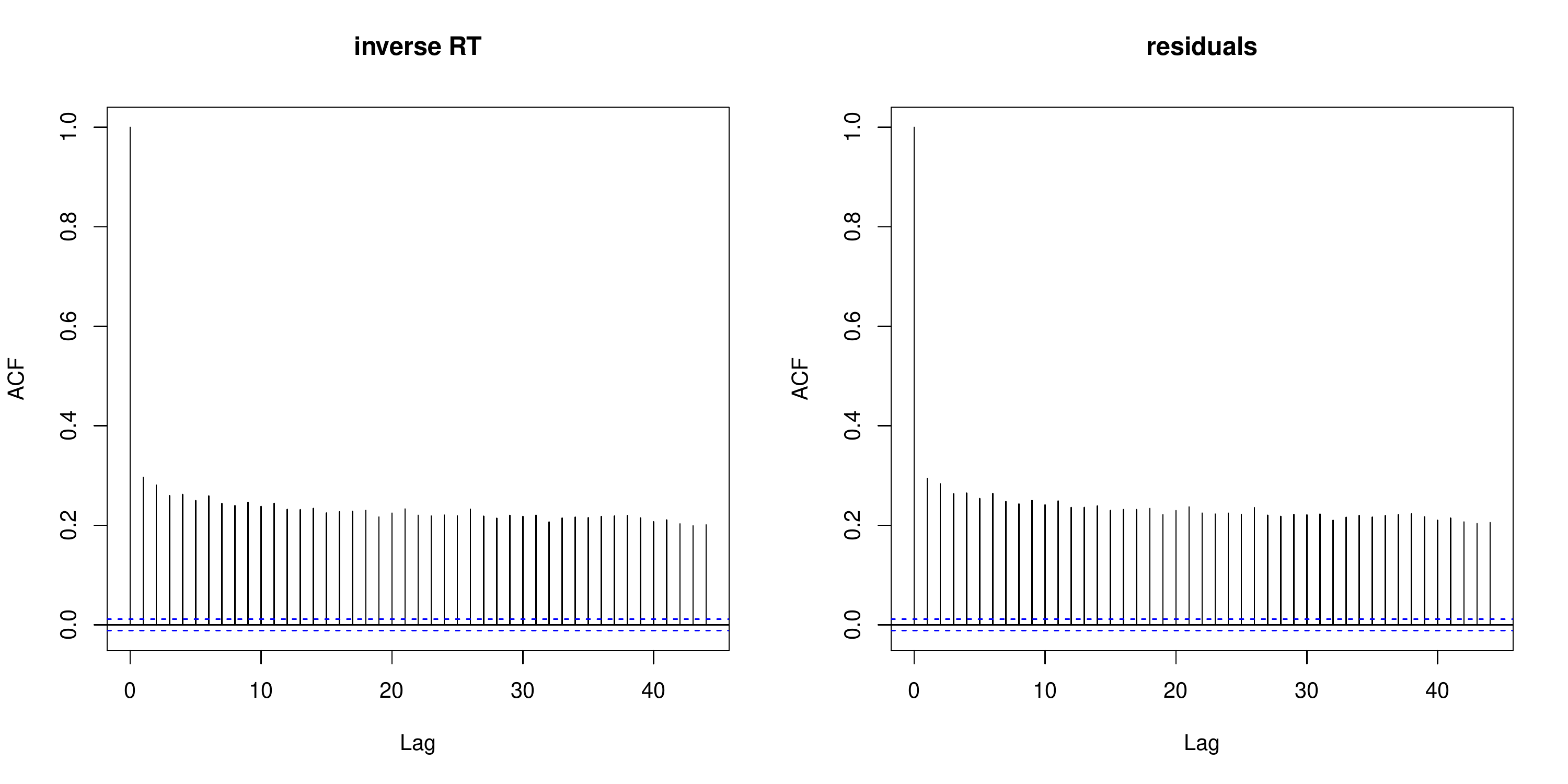}
\caption{
  Auto-correlation function of the inverse-transformed response latencies of
  subject 1 in the British Lexicon Project (left panel), and of the residuals
  of a model predicting inverse-transformed reaction time from lexicality (word
  vs. nonword).  Substantial autocorrelations are present even at lags of 40 trials.
}
\label{fig:acf2}
\end{figure}

\citet{barr2013random} proposed as gold standard for the analysis of experimental data with observations on combinations of subjects and items to fit maximally specified linear mixed effects models (LMMs). \citet{Bates:Kliegl:Vasishth:Baayen:2015} pointed out that such models run the risk of being overspecified, and \citet{Matuschek:2017} provided detailed discussion of the balance between power and type-I error in LMMs.  A study by \citet{Baayen:Vasishth:Kliegl:Bates:2017} raised another issue, namely that in psychometric data one often finds that sequences of response latencies observed over time as an experiment unfolds are not independent but are auto-correlated, often at substantial lags.  Auto-correlation functions (ACF) for subject 1 from the British Lexicon Project \citep{Keuleers:Lacey:Rastle:Brysbaert:2012} are presented in Figure~\ref{fig:acf2}.  The left panel presents the ACF for the inverse-transformed reaction times.  The right panel shows the ACF for the residuals of a model with {\tt lexicality} (word vs. nonword) as predictor.  In both cases, auto-correlations are markedly present even at lags of 40 trials.

The study by \citet{Baayen:Vasishth:Kliegl:Bates:2017} proposed to probe time-varying trends and temporal auto-correlations with the generalized additive mixed model (GAMM).  Introductions to the GAMM are provided in, e.g., \citet{Baayen:Vasishth:Kliegl:Bates:2017,wieling2018analyzing,Chuang:Fon:Papakyritsis:Baayen:2020}. \citet{Baayen:VanRij:DeCat:Wood:2018} provide detailed discussions of the kind of problems typically encountered when applying GAMMs to empirical data.  The question raised by \citet{BarrGam:2020} is whether time-dependent structure in the residuals of a LMM really affect the trustworthiness of the p-values produced by the LMM.  Central to the study of \citet{BarrGam:2020} are simulations for datasets implementing a design with two two-level treatments, one of which was within-subjects (henceforth $F_w$) and one of which was between-subjects (henceforth $F_b$).  They also contrasted a design in which the within-level factor was blocked with a design in which the within-level treatments were uniformly distributed over experimental time. 

For the modeling of time-varying trends in the data, \citet{BarrGam:2020} defined two kinds of functions of experimental time.  One function of time implemented a sine-wave.  The other function of time implemented an auto-correlative process.  The sine-wave function was implemented in two ways, once with the same phase but with an amplitude that varied from subject to subject, and once with the same amplitude but with a phase that varied by subject.  The auto-correlative process implemented either an AR(1) process or an AR(2) process.  When an AR(1) process is present in the errors, then the current error is a proportion $\rho$ of the preceding error augmented with Gaussian noise.  For an AR(2) process, the current error depends on the preceding two errors, again augmented with Gaussian noise.  In all, the simulations of \citet{BarrGam:2020} covered eight scenarios, the two types of sine waves without an AR(1) or AR(2) process, the two types of auto-correlative processes without sine waves, and the four combinations of sine waves and auto-correlative trends.

Analyses of the simulated data indicated no serious concerns about Type~I error rates. With respect to power, however, they observed the following.  Without blocking, GAMMs (as implemented in the {\bf mgcv} package for R, version $\leq$ 1.8-36) systematically outperformed LMMs by a small margin for the main effect of $F_w$ as well as for the interaction of $F_w$ by $F_b$. {\color{black} With blocking, GAMMs revealed substantially greater power for the LMMs for six of the scenarios, with respect to both the main effect of $F_w$ and the interaction, whereas for the remaining two scenarios, the LMMs showed greater power than the GAMMs by a small margin.} The power for the between-subject factor $F_b$ was remarkably different. For the varying amplitude, AR(1), and varying amplitude in combination with AR(1) scenarios, the GAMMs and LMMs showed the same power curves, {\color{black} with a slight loss of power} for the GAMMs when AR(1) noise was present. For the remaining scenarios, the power of the GAMMs was either substantially reduced or even completely eliminated, for both blocking and unblocking conditions.

{\color{black} The published version of \citet{BarrGam:2020}, \citet{thul2021using}, includes additional simulation studies that include time-varying effects that more closely approximate effects they observed for empirical datasets.  The conclusion the authors draw from these simulations is that, { \color{black}given proper randomization,} time-varying effects are orthogonal to the effects of the factors that are of theoretical interest, which leads them to question whether it is worth the effort to make use of GAMMs.} 

\citet{BarrGam:2020} offer little insight into why the GAMMs and the LMMs perform so differently. Their advice is not to use GAMMs {\color{black} for experiments with factorial designs}\footnote{\citet{thul2021using} acknowledge that GAMMs can be profitably used to analyse other kinds of data, such as data in which each observation constitutes a time series of its own, see \citet{Baayen:VanRij:DeCat:Wood:2018} and \citet{Chuang:Fon:Papakyritsis:Baayen:2020} for examples and discussion of such data.}, as these are ``complex, advanced techniques that are not fully understood'' and that have ``potential side-effects'' (p. 29){\color{black}, and the same point is made in \citet{thul2021using} {\color{black}(p. 14)}}.   As GAMMs are grounded in solid mathematics foundations  \citep{Hastie:Tibshirani:1990,Wood:2017}, we subjected the kind of data studied by \citet{BarrGam:2020} to closer scrutiny, comparing by means of simulation studies the performance of the LMMs and GAMMs.

However, it should be noted that the GAMM cannot model all the different datasets generated by \citet{BarrGam:2020}.   Whereas GAMMs can incorporate AR(1) noise in the errors, current implementations of the {\bf gam} and {\bf bam} functions in the {\bf mgcv} package do not allow incorporating AR(2) noise\footnote{{\color{black}It is possible to model AR(2) noise with the corARMA() function using the {\bf gamm} function in {\bf mgcv}. However, currently corARMA() does not work with factor smooths or by-smooths with linked smoothing parameters, the two smooths used to modify the time-varying effect in the current paper. Furthermore, {\bf gamm} depends on the {\bf nlme} package, which is highly restricted with respect to the random effects constellations, and although {\bf gamm4} builds on {\bf lme4}, which provides much faster algorithms, the latter does not provide handles for including autocorrelation processes in the errors.  Perhaps, interleaving with the MixedModels package in Julia will be possible in the future.}}.  In what follows, we therefore do not consider simulated datasets with AR(2) noise (see \citealt{Baayen:VanRij:DeCat:Wood:2018} for discussion of approximative strategies).  In our simulations, we also do not consider datasets with AR(1) noise, as the main problem reported for GAMMs  by \citet{BarrGam:2020} concerns the kind of smooths requested for modeling random wiggly curves, and not the presence or absence of AR(1) noise.

In the following sections, we first consider how time-varying processes generated with sine waves can be modeled, using on the one hand the linear mixed model and on the other hand the generalized additive mixed model.  We then proceed to consider random time-varying processes for which the functional form is not known.   We conclude this study with analyses of some empirical data with time-varying effects.

{\color{black} Before introducing our simulation studies, it is important to clarify the goal of these simulations. This goal is not to verify or even  prove that GAMMs are correct. This is impossible, as the kind of data to which GAMMs can be applied vary hugely, and no simulation study can ever cover all possible parameter settings of the infinite number of possible generating models.  The simulation studies presented below therefore serve a different goal.   In general, one goal of simulation studies can be to help especially non-mathematically trained analysts to develop a sense of the generating model. Another goal of simulations can be to gauge the power of experimental designs \citep[see, e.g.][]{Westfall:Kenny:Judd:2014}.  Yet another use of simulations is found in Bayesian modeling, where one can define generating models and use extensive (and unfortunately very carbon-hungry) simulations to sample from the posterior distribution of the parameters of interest.\footnote{{\color{black}Readers interested in integrating GAMMs with JAGS are referred to \citet{Wood:2016jagam}}.}  Our simulations do not fall into this class of models, as the datasets of interest to us can be analyzed straightforwardly with the (carbon-lean) empirical Bayes methods underpinning GAMMs. 

The goal of the simulations presented below is much simpler. The study of \citet{thul2021using} reports a serious problem for GAMMs given a particular experimental design, and we therefore set ourselves the task of replicating this problem, for both identical and very different parameter settings, with the aim of clarifying whether it is indeed the mathematics of GAMMs that are poorly understood within the community of mathematical and computational statistics, as claimed by \citet{thul2021using}, or whether \citet{thul2021using} discovered a bug in the GAMM implementation provided by {\bf mgcv}.} 

\section{Sine waves}

Following \citet{BarrGam:2020}, we consider time-varying effects that take the shape of sine waves.  We begin with sine waves that vary only in amplitude but not with respect to their phase,  in combination with a two-treatment design that includes a between-subject treatment $F_b$ and a within-subject treatment $F_w$.  The within-subject treatment is blocked, such that a given subject receives one treatment level (e.g., A) in the first half of the trials, and the other treatment level (B) in the second half of the trials, or vice versa.

\subsection{Sine waves with varying amplitudes}\label{sec:amp}

\begin{figure}[htbp]
    \centering
    \includegraphics[width=0.46\textwidth]{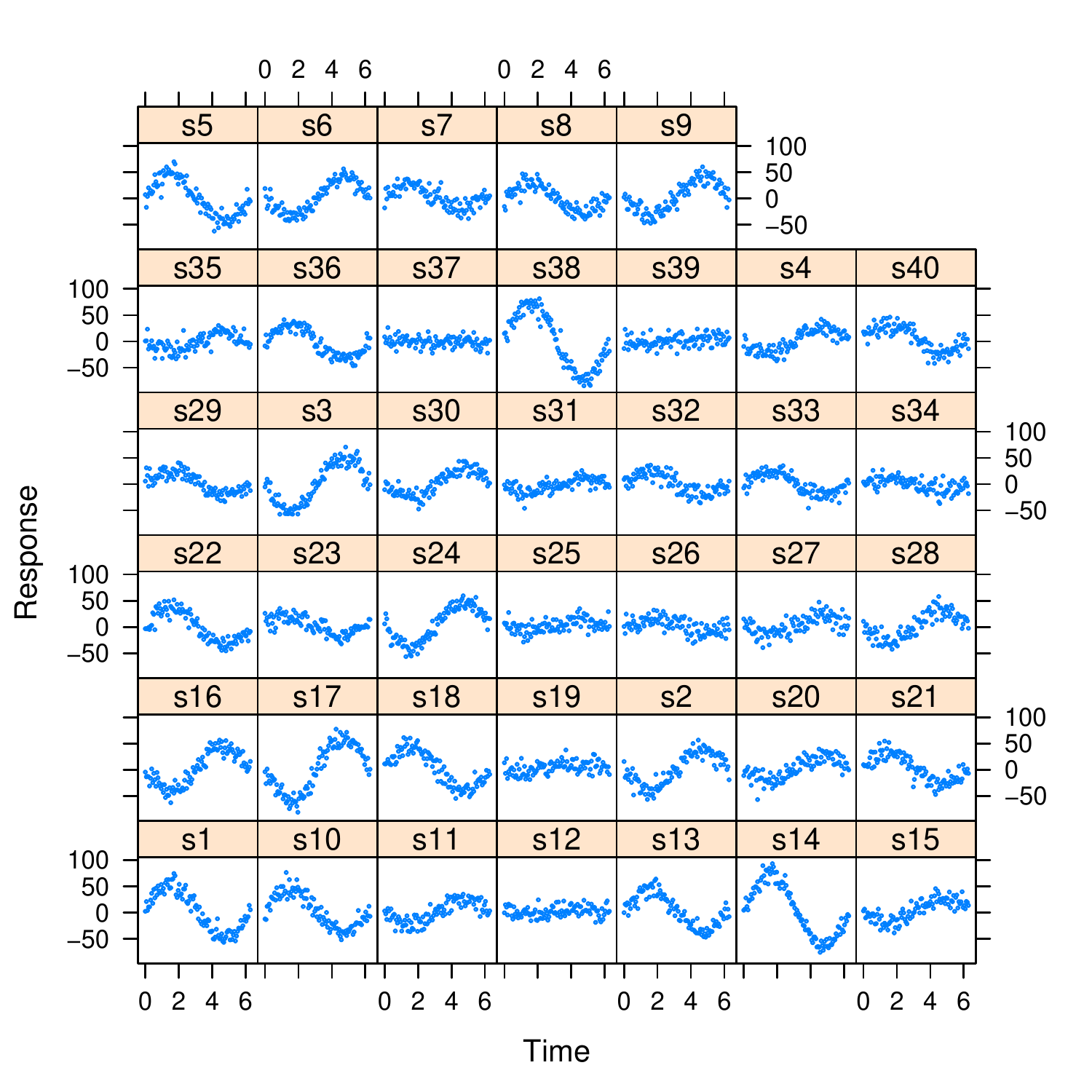}\includegraphics[width=0.46\textwidth]{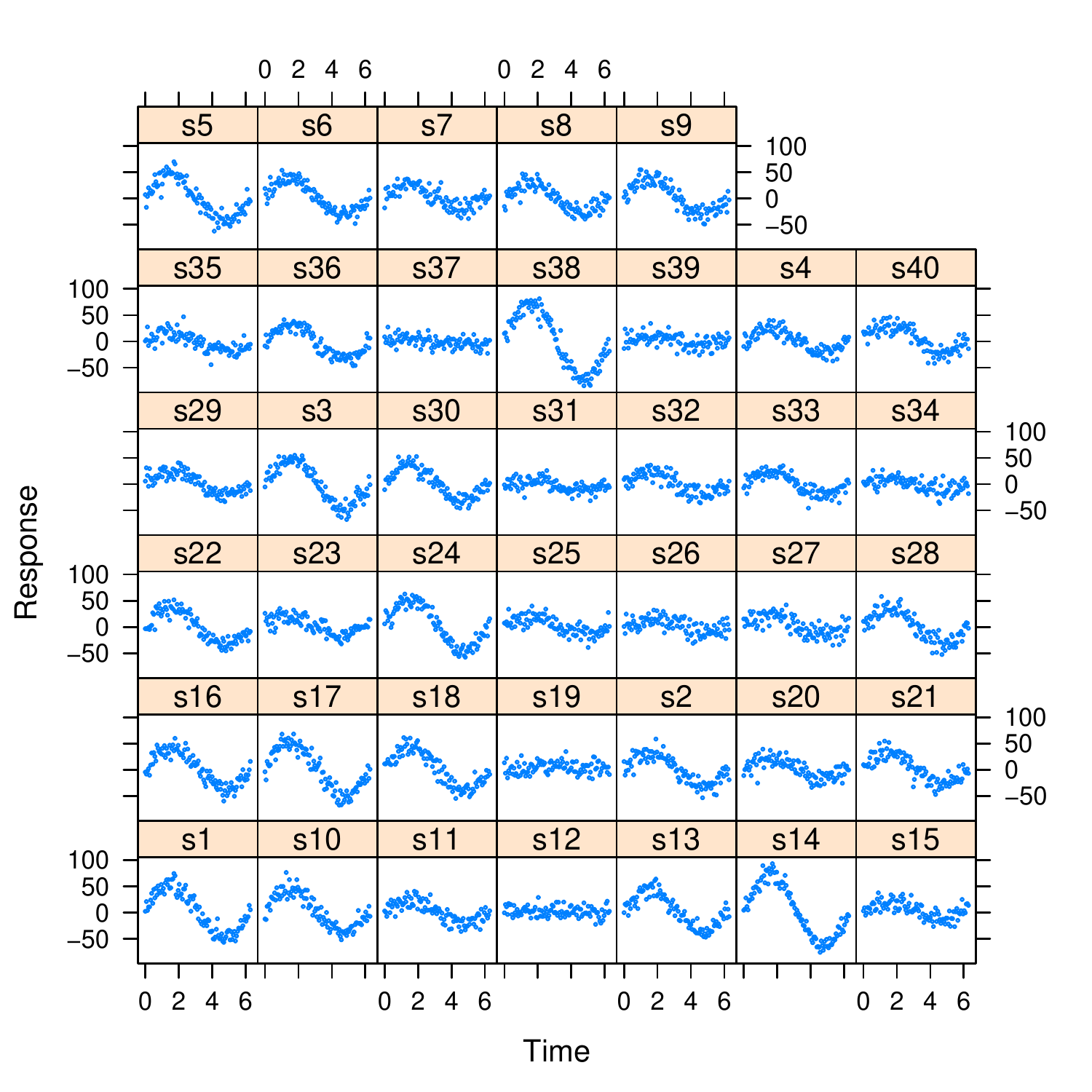}
    \caption{The sine waves for the individual subjects in the {\tt amp} dataset (left) and the {\tt ampabs} dataset (right).}
    \label{fig:amp}
\end{figure}

The response for subject $i$ at time $t$, $y_{ti}$,  
\begin{equation}
y_{ti} = \beta + b_{i} +  \beta_{W_t}  + \beta_{B_{i}} + \alpha_i \sin(t) + \varepsilon_{ti},
\label{eq:amp}
\end{equation}
is the sum of a general intercept $\beta$, by-subject random intercepts $b_{i}$, the within-subject treatment $\beta_{W_t}$ at time $t$ (factor levels A, B with A as reference level), the between-subject treatment $\beta_{B_{i}}$ (factor levels X, Y with X as reference level), and the sine wave $\sin(t)$, which has subject-specific amplitude $\alpha_i$.  To complete the model specification, the by-subject random intercepts,  the by-subject amplitudes, and the error are all defined to follow independent normal distributions:
\begin{eqnarray*}
b_{i} & \sim & {\cal N}(0, \sigma_b), \\
\alpha_i & \sim & {\cal N}(0, \sigma_{\alpha}),\\
\varepsilon_{ti} & \sim & {\cal N}(0, \sigma).
\end{eqnarray*}
A dataset simulated according to these specifications (100 timesteps in the interval $[0, 2\pi]$, 40 subjects, $\beta = 0, \beta_{w} = 2, \beta_{b} = 2$, $\sigma_{b} = 1$, $\sigma_{\alpha} = 32$, $\sigma = 10)$ is {\tt amp}, available in the supplementary materials at \url{https://osf.io/fbndc/}.\footnote{{\color{black}We generated and made available these datasets to enable comparison for exactly the same data points of the LMM using the implementations of the {\bf lme4} package in R and the {\bf MixedModels} package in Julia.}}  The left panel of Figure~\ref{fig:amp} visualizes the sine waves for each of the subjects. 

Conveniently, even though a model term like $(1+\sin(t)|\text{subject})$ is nonlinear in $t$, it still just creates a part of the $\bm{Z}$ matrix of the model 
$$
\mbox{E}[\bm{y}|\bm{B} = \bm{b}] = \bm{X}\bm{\beta} + \bm{Z}\bm{b},
$$
\noindent
which is fixed once the data and model formula are known.  Hence, the models is linear in $\bm \beta$ and $\bm b$ and can therefore be fit using the {\tt lmer} function from the {\bf lme4} package for R, as follows:\footnote{We are indebted to Douglas Bates for pointing this out to us.}

\begin{verbatim}
lmer(Response ~ Factor_Within + Factor_Between +  
                (1|Subject) + (0 + sin(Time)|Subject),
     data = amp)
\end{verbatim}
Ideally, one would want to fit a model with a population sine wave, as in general one would want any covariate in the random effects to be also present in the fixed effects.  But as no such wave is present in the generating model, we leave this population sine out of the model specification. Table~\ref{tab:amp} presents the estimates of the model parameters.  The true values of the treatment coefficients are all within two standard errors of the estimated values, and the estimated standard deviations are all quite close to the values used to generate the data. 

\begin{table}[hbp]
\centering
\caption{Coefficients of the fixed effects of the LMM fitted to the {\bf amp dataset}. The parameters of the generating model are $\beta = 0, \beta_w = 2, \beta_b = 2, \sigma = 10, \sigma_b = 1, \sigma_\alpha= 32$.} \vspace*{0.6\baselineskip}
\label{tab:amp}
\begin{tabular}{lrrr} \hline
                                       & Estimate   & Std. Error &  t value  \\ \hline
Intercept                              &  -0.6562   &     0.4738 &   -1.385  \\
Within-Subject Treatment = B           &   2.9291   &     0.7131 &    4.108  \\
Between-Subject Treatment = Y          &   2.3630   &     0.4413 &    5.354  \\ \hline
\end{tabular} 
\ \\ \vspace*{0.4\baselineskip}
$\hat{\sigma} = 10.0617, \hat{\sigma}_b = 0.9671, \hat{\sigma}_\alpha = 33.8683$ 
\end{table}

We can impose a constraint on the sine waves such that the positive inflection of the sine wave always coincides with the first half of the simulated experiment, mirroring the set-up of \citet{BarrGam:2020} (scenario  2 as visualized in their Figure 5).  To do so, we sampled amplitudes from a normal distribution, but used the absolute values of these amplitudes.  For the present parameters (40 subjects, $\sigma_\alpha = 32$), this leads to only mild departures from normality. The sine waves in the resulting dataset, available as {\tt ampabs} in the supplementary materials, is visualized in the right panel of Figure~\ref{fig:amp}.  

The LMM that recaptures the generating parameters fairly well has exactly the same model specification as for the model fitted to the data with unconstrained amplitudes. The estimated parameters are presented in the center column of Table~\ref{tab:amp_comp}. Again, all estimates are good.   However, when the sine wave is dropped from the random effects structure, 
\begin{verbatim}
lmer(Response ~ Factor_Within + Factor_Between + (1|Subject), 
     data=ampabs)
\end{verbatim}
the model fails to reconstruct the contrast for the within-subject treatment, as shown in the third column of Table~\ref{tab:amp_comp}. The coefficient for $\beta_w$ is twice as large as it should be, with the wrong sign.  This is because the sine waves happen to have negative values predominantly for the second level of the within-subject treatment factor (mean -7.1, $p < 0.0001$ according to a mixed model regressing the sine wave component on the two treatment factors). Furthermore, the standard deviation of the residual error is much too high, and the by-subject random intercepts are not recovered at all.  {\color{black} It is also noteworthy that the standard deviations of the fixed effects are larger in the LMM that ignores the sine wave (0.81 and 0.81 for $\beta_w$ and $\beta_b$) compared to the LMM that includes the sine wave (0.71 and 0.44 for $\beta_w$ and $\beta_b$ respectively).} In short, ignoring  temporal effects in experimental data with a blocked design can give rise to incorrect and misleading estimates of effect sizes for a given dataset.   

\begin{table}[htbp]
\centering
\caption{Parameters of the generating model and corresponding estimates based on a LMM that includes a sine wave, and a LMM that ignores the sine wave ({\bf ampabs} dataset). The amplitudes of the sine wave are sampled from a normal distribution, but their absolute value is taken to ensure that the positive inflection of the sine wave always coincides with the first half of the experiment (see Figure~\ref{fig:amp}, right panel).} \vspace{0.6\baselineskip}
\label{tab:amp_comp}
\begin{tabular}{lrrr} \hline
                & generating model  & LMM (sine wave modeled)    &  LMM (sine wave ignored)       \\ \hline
$\beta$         &   0               &     -0.61                   &      3.0                       \\ 
$\beta_w$       &   2               &      2.83                   &     -4.39                      \\
$\beta_b$       &   2               &      2.36                   &      2.36                      \\
$\sigma_b$      &   1               &      0.97                   &      0.00                      \\
$\sigma_\alpha$ &  32               &     33.74                   &       -                        \\
$\sigma$        &  10               &     10.06                   &     25.56                      \\ \hline
log likelihood  &                   & -15048.7                    & -18636.9                       \\
AIC             &                   &  30109.4                    &  37283.8                       \\ \hline
\end{tabular}
\end{table}

How does the generalized additive model fare with this kind of data?  The GAMM toolkit as implemented in the {\bf mgcv} package for {\tt R} \citep{Wood:2017} makes available many different kinds of smoothing splines.  For the modeling of by-subject wiggly curves, the non-linear counterpart of by-subject regression lines with individual slopes and intercepts, factor smooths, often are a natural choice.   Factor smooths are constructed in such a way that the smooths for individual subjects all share the same smoothing parameter $\lambda$. This smoothing parameter regulates the balance of staying faithful to the data (minimizing the sum of squares) and keeping the model simple (by penalizing wiggliness).  In this way, the factor smooth implements the idea that subjects, coming from the same population, should be similar with respect their wiggliness, and hence identical with respect to the amount of penalization that their smooths require. Furthermore, for wiggly random effects, factor smooths are usually specified in such a way that even the contributions of the linear basis function can be penalized.  Finally, factor smooths are optimized computationally for modeling with the {\tt gamm} and {\tt gamm4} functions.   In actual modeling, we can distinguish three situations.
\begin{enumerate}
    \item There is no functional dependency of the response variable on the pertinent covariate for any subject.  In this case, the factor smooth will return random intercepts, which in partial effect plots with time as predictor will show up as horizontal lines. 
    \item The functional dependency of the response variable on the covariate is linear, and could have been modeled with by-subject random intercepts and by-subject random slopes.  In this case, the GAMM will penalize all non-linear basis functions, while retaining the linear basis function.  In partial effect plots, these random effects will show up as straight lines with non-zero slope.
    \item The functional dependency of the response variable on the covariate is truly wiggly in nature.  In this case, the GAMM will return a separate unique wiggly curve for each subject.
\end{enumerate}
Because factor smooths already incorporate by-subject random intercepts, the model specification should not request separate by-subject random intercepts.  A GAMM for the {\tt ampabs} dataset with a factor smooth, requesting 20 basis function (k=20), is specified as follows:
\begin{verbatim}
bam(Response ~ Factor_Within + Factor_Between +  
               s(Time, Subject, bs="fs", m=1, k=20), 
    data = ampabs)
\end{verbatim}

Table~\ref{tab:ampabs_fs_correct} presents the model summary obtained with {\bf mgcv} version 1.8.36 or higher. {\color{black} In this table, `edf' denotes the effective degrees of freedom (defined as the sum of the shrinkage proportions of the weights of the basis functions of a smooth) and. {\color{black} `Ref.df' refers to the reference degrees of freedom \citep{Wood:2013a} that are used in computing the test statistic and its p-values, but as the null distributions are non-standard, the Ref.df are not very interpretable.  We list them for completeness only.}} The estimated variance components of this model are obtained with {\tt gam.vcomp}, and are presented in Table~\ref{tab:ampabs_fs_varcomp}. This model does not pick up the effect of the within-subject treatment, but the magnitude of this treatment effect is estimated correctly, and the estimates of $\sigma$ and $\sigma_b$ are precise.

\begin{table}[ht]
\centering
\caption{Model summary for a GAMM with a factor smooth fitted to the {\bf ampabs} dataset {\color{black}($\beta = 0, \beta_w = 2, \beta_b = 2, \sigma_b = 1, \sigma_\alpha = 32, \sigma = 10$)}. fREML: 15475; AIC: 30212.13.}  
\label{tab:ampabs_fs_correct}
\begin{tabular}{lrrrr}
   \hline
A. parametric coefficients & Estimate & Std. Error & t-value & p-value \\ 
  Intercept & -0.3429 & 0.7864 & -0.4360 & 0.6629 \\ 
  Factor\_WithinB & 2.3024 & 1.4437 & 1.5948 & 0.1109 \\ 
  Factor\_BetweenY & 2.3630 & 0.4413 & 5.3547 & $<$ 0.0001 \\ 
   \hline
B. smooth terms & edf & Ref.df & F-value & p-value \\ 
  factor smooth Time $\times$ Subject & 532.6603 & 798.0000 & 28.2142 & $<$ 0.0001 \\ 
   \hline
\end{tabular}
\end{table}

\begin{table}[htbp]
\centering
\caption{Estimated standard deviations with 95\% confidence interval for the variance components for the GAMM fitted to the {\bf ampabs} dataset, using a factor smooth. $\sigma_b$ is the standard deviation of the by-subject random intercepts. $\sigma_t^2$ is the prior variance of the coefficients of the by-subject smooths.}
\label{tab:ampabs_fs_varcomp}
\begin{tabular}{lrrr} \hline
                               &  std.dev  & lower  &  upper    \\ \hline
s(Time,Subject)1 ($\sigma_t$)  &    17.42  & 16.38  & 18.53     \\
s(Time,Subject)2 ($\sigma_b$)  &     0.98  &  0.62  &  1.55     \\
scale  ($\sigma$)              &     9.92  &  9.69  & 10.16     \\ \hline
\end{tabular}
\end{table}

The data can also be fitted using a by-smooth.  Similarly to factor smooths, by-smooths estimate a different wiggly curve for each level of a given factor. If the default arguments for specifying such smooths are used, {\bf mgcv} constructs by-smooths formed by a fixed effect and a smooth effect for each subject. The fixed effect is a subject-specific slope, while the smooth part is constructed by using the same thin-plate splines basis functions for all subjects.  
{\color{black} While the wiggliness of each subject-specific effect is quantified by the same smoothing penalty (typically, the integrated squared curvature), each subject has its own smoothing parameter. Given that the latter quantifies the strength of the penalty, subject-specific effects can differ substantially with respect to the amount of wiggliness.} In order to model subjects as a random sample from a population characterized by similar degrees of wiggliness, we use the {\tt id = 1} directive to enforce that the curves for the subjects will all be estimated with the same smoothing parameter $\lambda$. To make the model directly comparable to the one we built above using factor smooths, we also set the parameter $m$ to 1 to allow penalization of the linear basis function as well. As the basis functions of by-smooths do not include an intercept (the splines basis is orthogonal to it), by-subject random intercepts are specified separately.
\begin{verbatim}
bam(Response ~ Factor_Within + Factor_Between +  
               s(Subject, bs="re") + s(Time, by = Subject, id = 1, m = 1, k = 20),
    data = ampabs)
\end{verbatim}
Table~\ref{tab:ampabs_by} presents the summary of this model.  The estimates for the treatment effects and their standard errors are exactly the same as for the model fitted with a factor smooth.  Furthermore, the estimates of the variance components are also identical to those presented in Table~\ref{tab:ampabs_fs_varcomp}.  An advantage offered by the by-smooths is that the summary reports, for each subject, whether there is a significant time-varying effect for that subject.  The disadvantage of using by-smooths as compared to factor smooths is that factor smooths are set up in such a way that they can be estimated more efficiently when the model is fitted via the {\tt gamm} or {\tt gamm4} functions, which can be important when working with random effects with large numbers of levels.

\begin{table}[t]
\centering
\caption{Model summary for a GAMM with a `by' smooth fitted to the {\bf ampabs} dataset {\color{black}($\beta = 0, \beta_w = 2, \beta_b = 2, \sigma_b = 1, \sigma_\alpha = 32, \sigma = 10$)}. fREML: 15475; AIC: 30212.13; $\hat{\sigma}_b = 0.98$; $\hat{\sigma} = 9.92$;  Of the 40 by-subject smooths, only four are shown.}  \vspace*{0.6\baselineskip}
\label{tab:ampabs_by}
\begin{tabular}{lrrrr} \hline
A. parametric coefficients & Estimate & Std. Error & t-value & p-value \\ 
  Intercept & -0.3429 & 0.7864 & -0.4360 & 0.6629 \\ 
  Factor\_WithinB & 2.3024 & 1.4437 & 1.5948 & 0.1109 \\ 
  Factor\_BetweenY & 2.3630 & 0.4413 & 5.3547 & $<$ 0.0001 \\ 
   \hline
B. smooth terms & edf & Ref.df & F-value & p-value \\ 
  s(Subject) & 18.7803 & 38.0000 & 0.9771 & 0.0003 \\ 
  s(Time):Subjects1 & 12.8470 & 19.0000 & 51.0953 & $<$ 0.0001 \\ 
  s(Time):Subjects10 & 12.8470 & 19.0000 & 34.2018 & $<$ 0.0001 \\ 
  s(Time):Subjects11 & 12.8470 & 19.0000 & 10.2052 & $<$ 0.0001 \\ 
  s(Time):Subjects12 & 12.8470 & 19.0000 & 0.8252 & 0.2242 \\ 
  \ldots \\ \hline
\end{tabular}
\end{table}

Unfortunately, versions of the {\bf mgcv} package prior to version 1.8-36 contained a bug in how the factor smooths were set up.  Table~\ref{tab:ampabs_fs} illustrates that with the older version (1.8-31), the effect of the between-factor is overestimated, and Table~\ref{tab:ampabs_fs_varcomp_old} clarifies that the estimate of $\sigma_b$ is incorrectly estimated to be close to zero, with an extremely wide confidence interval.  It turns out that in the implementation of the factor smooth, the by-subject intercept of the factor smooths was not made orthogonal to the  by-subject smooth part, in contrast to the orthogonality enforced in by-smooths. In {\bf mgcv} version 1.8-36, orthogonality is now imposed also for factor smooths, which solves the problem.

\begin{table}[ht]
\centering
\caption{Model summary for a GAMM with a factor smooth fitted to the {\bf ampabs} dataset {\color{black}($\beta = 0, \beta_w = 2, \beta_b = 2, \sigma_b = 1, \sigma_\alpha = 32, \sigma = 10$)}, using an older version of {\bf mgcv}. $\hat{\sigma} = 9.94$; fREML: 15511; AIC: 30213.59.}  \vspace*{0.6\baselineskip}
\label{tab:ampabs_fs}
\begin{tabular}{lrrrr} \hline
A. parametric coefficients   & Estimate & Std. Error & t-value & p-value \\ 
  Intercept        & -0.8714 & 1.0735   & -0.8117 & 0.4170 \\ 
  Factor\_WithinB  &  2.2546 & 1.4367   &  1.5693 & 0.1167 \\ 
  Factor\_BetweenY &  3.6016 & 1.1282   &  3.1924 & 0.0014 \\ \hline
B. smooth terms    & edf     & Ref.df   & F-value & p-value \\ 
  factor smooth Time $\times$ Subject   & 519.0138 & 798.0000 & 28.0892 & $<$ 0.0001 \\ \hline
\end{tabular}
\end{table}

\begin{table}[ht]
\centering
\caption{Estimated standard deviations with 95\% confidence interval for the variance components for the GAMM fitted to the {\bf ampabs} dataset {\color{black}($\sigma_b = 1, \sigma = 10$)} with a factor smooth, using an older version of {\bf mgcv}. {\color{black} The standard deviation for the amplitude of the sine waves was $\sigma_\alpha = 32$; however, the parameter $\sigma_t$ reported in the present table is the prior variance of the coefficients of the by-subject smooths, i.e., $\sigma^2/\lambda$ in $\bm{\beta} \sim {\cal N}(0, \mbox{S}^{-} \sigma^2/\lambda)$, see \citet{Wood:2017} for further details. Note the immensely wide confidence interval for $\sigma_b$}.}  \vspace*{0.6\baselineskip}
\label{tab:ampabs_fs_varcomp_old}
\begin{tabular}{lrrrr} \hline
                               &    std.dev &        lower &        upper  \\ \hline
s(Time,Subject)1  ($\sigma_t$)             & 17.1873388 & 1.616096e+01 & 1.827890e+01  \\
s(Time,Subject)2 ($\sigma_b$)  &  0.0170401 & 6.914402e-125& 4.199425e+120 \\
scale  ($\sigma$)              &  9.9430820 & 9.711725e+00 & 1.017995e+01  \\ \hline
\end{tabular}
\end{table}

In the remainder of this study, we document that with the corrected factor smooths, the kind of problems documented by \citet{BarrGam:2020} for various kinds of simulated data no longer occur.

\subsection{Sine waves with varying phase}\label{sec:phase}

Thus far, we have considered datasets with sine waves that all share the same phase, and that all have their intercept at the origin.  Following \citet{BarrGam:2020}, we next consider simulated data with by-subject sine waves that all have the same amplitude, but that differ with respect to their phase $\phi$.  The dataset {\tt phase}, available in the supplementary materials, implements such time-varying sine waves, using the following generating model:   
\begin{equation}
y_{ti} = \beta + b_{i} +  \beta_{W_t}  + \beta_{B_{i}} + \alpha \sin(t-\phi_i) + \varepsilon_{ti},
\label{eq:phase}
\end{equation}
with
\begin{eqnarray*}
b_{i} & \sim & {\cal N}(0, \sigma_{b}), \\
\phi_i & \sim & {\cal N}(0, \sigma_{\phi}),\\
\varepsilon_{ti} & \sim & {\cal N}(0, \sigma),
\end{eqnarray*}
with zero covariances. The parameters that generated this dataset are $\beta=0,  \beta_{b} = 2, \beta_w = 2, \alpha = 8, \sigma_b = 1, \sigma_\phi = 2, \sigma=8$. For this dataset, the within-subject treatment is not blocked.  

The challenge that this dataset poses to the analyst is that, in addition to by-subject random intercepts and the effects of the two treatments, the phase of the sine wave also contributes to subjects' intercepts. Here, one could in principle use the non-linear mixed model using the {\bf nlme} package \citep{Pinheiro:Bates:2000}, but attempts to do so indicated the model was too unstable to be automatically fitted to large numbers of simulated datasets.   We therefore focus on how the generalized additive mixed model performs, using factor smooths.
\begin{verbatim}
bam(Response ~ Factor_Within + Factor_Between +  
               s(Time, Subject, bs = "fs", k = 20, m = 1),
    data=phase)
\end{verbatim}
Table~\ref{tab:phase} provides the summary of the model, and Table~\ref{tab:phase_fs_varcomp} presents the variance components.  

The model succeeds in detecting the two treatment effects, with estimates that contain the true values within their 95\% confidence intervals. The estimates of the variance components $\sigma_b$ and $\sigma$ are also very close to the true values.   The exact same results are obtained with a properly specified by-smooths, details are provided in the supplementary materials.


\begin{table}[htbp]
\centering
\caption{Model summary for a GAMM with a factor smooth fitted to the {\bf phase} dataset {\color{black}($\beta=0,  \beta_{b} = 2, \beta_w = 2, \alpha = 8, \sigma_b = 1, \sigma_\phi = 2, \sigma=8$)}. fREML: 14216; AIC: 28162.59; $\hat{\sigma}_b = 1.07$; $\hat{\sigma} = 7.87$.}  \vspace*{0.6\baselineskip}
\label{tab:phase}
\begin{tabular}{lrrrr} \hline
A. parametric coefficients & Estimate & Std. Error & t-value & p-value \\ 
  Intercept & -0.1883 & 0.3232 & -0.5827 & 0.5601 \\ 
  Factor\_WithinB & 1.6017 & 0.2583 & 6.2022 & $<$ 0.0001 \\ 
  Factor\_BetweenY & 2.4322 & 0.4189 & 5.8057 & $<$ 0.0001 \\ \hline
B. smooth terms & edf & Ref.df & F-value & p-value \\ 
  s(Time,Subject) & 307.9472 & 798.0000 & 2.8042 & $<$ 0.0001 \\ \hline
\end{tabular}
\end{table}

In summary, when analysing datasets with between-subject treatments or blocked within-subject treatments, it is necessary to uncouple the estimation of by-subject intercepts from by-subject wiggly curves.   This has always been possible with the {\bf mgcv} package by using `by' smooths, constrained to have the same smoothing parameter, and for full equivalence with the factor smooth that we made use of, by setting $m=1$. As of {\bf mgcv} version 1.8-36 such uncoupling is performed also under factor smooths, which are more efficient when the model is fitted with {\tt gamm} or {\tt gamm4}. 

\begin{table}[ht]
\centering
\caption{Estimated standard deviations with 95\% confidence interval for the variance components for the GAMM fitted to the {\bf phase} dataset with a factor smooth.}  \vspace*{0.6\baselineskip}
\label{tab:phase_fs_varcomp}
\begin{tabular}{lrrrr} \hline
                               &    std.dev &        lower &        upper  \\ \hline
s(Time,Subject)1  ($\sigma_t$) &  5.8174 & 5.2902 & 6.3971  \\
s(Time,Subject)2 ($\sigma_b$)  &  1.0654 & 0.7524 & 1.5086 \\
scale  ($\sigma$)              &  7.8741 & 7.6939 & 8.0585  \\ \hline
\end{tabular}
\end{table}

\section{Power and Type-I error rate}

Thus far, we have investigated individual simulated datasets, focusing on the quality of the parameter estimates.  In what follows, we inspect power and Type-I error rates. 

\subsection{Sine waves with varying absolute amplitude} 

We evaluated performance of the LMM,  the LMM with a sine random effect, henceforth LMMsine, and GAMMs with factor smooths and `by' smooths, on the basis of 1000 simulated datasets. The generating model was identical to that given in (\ref{eq:amp}), with the same parameter values except for $\beta_b$ (set to 1) and for $\sigma_{\alpha}$, which was lowered from 32 to 8.  {\color{black} By reducing $\sigma_{\alpha}$, the datasets were not almost completely dominated by the sine waves, as a consequence, the probability of observing false positives increased.}  The within-subject treatment was blocked, and negative amplitudes were converted into positive ones.   By-subject random slopes for the within-subject treatment were not included in the generating model. We fitted five models to each simulated dataset:
\begin{enumerate}
    \item LMMsine: a linear mixed model with a sine random effect;
    \item LMMmin: a linear mixed model that ignores experimental time;
    \item LMMmax: a maximal linear mixed model that ignores experimental time, and that includes by-subject random slopes for the within-subject treatment, following the recommendations of \citet{barr2013random}, even though such random slopes are not part of the generating model;
    \item GAMMfs: a GAMM with a factor smooth for experimental time;
    \item GAMMby: a GAMM with a `by' smooth for experimental time, this model did not include by-subject random slopes for the within-subject treatment as such slopes were not part of the generating model.
\end{enumerate}
Table~\ref{tab:ampabs1000} provides a summary of the results obtained.  The LMMsine  provides precise estimates for all fixed-effect parameters, as well as for the standard deviations, including the standard deviation of the random sine wave amplitudes.  {\color{black} Error rates are nominal, i.e., in accordance with the alpha-level.}  Power for the between-subject treatment is smaller than that for the within-subject treatment.

GAMMs with factor smooths or `by' smooths have reduced power compared to the LMMsine for the within-subject treatment, but power for the between-subject treatment is comparable to that of the LMMsine, and error rates are again slightly above nominal.   

The minimal linear mixed model (LMMmin) has the greatest power of all models for the within-subject treatment, but here it has an unacceptably high error rate:  LMMmin does not survive the confound of the within-subject effect and the sine waves.  Power for the between-subject effect is comparable to that of the LMMsine. 

A maximal LMM (LMMmax) has catastrophically low power when it comes to detecting the effect of the within-subject treatment.  Because by-subject random slopes for the within-subject factor are included, even though random slopes are not part of the ground truth, the random intercepts absorb the effect of the sine wave, making it impossible for the model to detect the within-subject treatment effect.  Unsurprisingly, as can be seen in Table~\ref{tab:ampabs1000}, the maximal LMM substantially overestimates the standard deviation $\sigma_b$ for the by-subject random intercepts.\footnote{{\color{black}
For some configurations of parameters, the number of simulation runs for which a significant effect at $\alpha=0.01$ was observed was less than 10 out of 1000.  This happens for the intercept of models LMMsine and LMMmax.  However, since in all simulations, the ground truth for the intercept is zero, power and type~I error are identical, and this holds across all configurations of parameters.  Extremely low power also occurs when a model is misspecified: in this case, the models fitted to the data are different from the mechanism generating the data.  Biased tests are a straightforward consequence. This explains why, for example, LMMmax only detects within-subject treatments a mere 18 out of 1000 times.
}}

A final observation concerns the variances of the estimated coefficients, listed in the rightmost subtable of  Table~\ref{tab:ampabs1000}.  These variances are smallest for the correct model, LMMsine, followed by the GAMMs, and largest for the misspecified models (LMMmin and LMMmax).  Thus, a misspecified model may yield, for a particular dataset, incorrect and misleading estimates, even though in the long run, across a thousand experiments/simulation runs, the mean coefficients are fine.  Above, we encountered such an example for the {\tt absamp} dataset (see Table~\ref{tab:amp_comp}, which lists an LMM $\beta_w$ estimated at -4.39 ($t = -5.4$), instead of at 2).  In other words, {\color{black} this example illustrates that by ignoring time-varying trends in the data, obtaining consistent results across replication experiments can be difficult.} 

The analyst's best choice that emerges from Table~\ref{tab:ampabs1000} is the LMMsine.  Not only does this model provide excellent estimates of the by-subject variability in amplitudes,  it also offers the best power for the within-subject treatment and comparable power for the between-subject treatment.  That the LMMsine is preferable over the GAMMs smooths is unsurprising. The LMMsine is given the functional form of the time-varying effect beforehand, a sine wave, whereas the GAMM has to reconstruct this functional form from the data. 

\subsection{Sine waves with varying phase} 

Next consider power and Type~I error rate for sine waves with varying phase and identical amplitude. The generating model was based on the formula given in (\ref{eq:phase}), with three modifications. First, we included the interaction effect $F_b$ by $F_w$ ($\beta_{bw} = -3$). Second, we increased $\sigma$ to 16 to make it more difficult for the models to detect the treatment effects. Finally, an additional error term for by-subject random slopes was added, with standard deviation ($\sigma_{b_w}$) set to 1. Similar to the simulation presented in Section \ref{sec:phase}, the within-subject treatment was not blocked.

\begin{sidewaystable}[p]
\centering
\caption{
{\bf Sine waves with varying absolute amplitude}: Mean estimated parameters and number of simulation runs out of a total of 1000 in which a treatment effect is significant at $\alpha = 0.01$. In the generating model, $\beta = 0, \beta_w = 2, \beta_b = 1, \sigma_\alpha = 8, \sigma_b = 1, \sigma = 10$.  $\sigma_t$ pertains to the wiggliness of the by-subject time-varying effects. For the simulations for the Type~I Error rate, $\beta_w = \beta_b = 0$.  LMMsine's estimated standard deviation of the amplitude: $\hat{\sigma}_\alpha = 7.96$ for both power and type I error simulations.  Variances: the variances corresponding to the means of the fixed-effect parameters.} \vspace*{0.7\baselineskip}
\label{tab:ampabs1000}
{\footnotesize
\begin{tabular}{|l|rrrrrr|rrr|rr|} \hline
& \multicolumn{6}{c|}{mean estimated parameters and power} & \multicolumn{3}{c|}{estimated std. deviations} & \multicolumn{2}{c|}{variances of estimated coefficients} \\ \hline
        & $\beta$ & n($\beta$) & $\beta_w$ & n($\beta_w$) & $\beta_b$  & n($\beta_b$)  &     $\sigma$ &  $\sigma_b$ &  $\sigma_t$       &   VAR[$\beta_w$] & VAR[$\beta_b$] \\ \hline
LMMsine & 0.0132  &          8 & 1.9915    &      658     &  0.9812    &        380    &     9.9984   &   0.9696    &                   &   0.4211    &    0.1981 \\
GAMMfs  & 0.0215  &         13 & 1.9750    &      423     &  0.9812    &        379    &     9.9770   &   0.9722    &            5.6891 &   0.6328    &    0.1981 \\
GAMMby  & 0.0215  &         13 & 1.9750    &      423     &  0.9812    &        379    &     9.9770   &   0.9722    &            5.6891 &   0.6328    &    0.1981 \\
LMMmin  & 0.0263  &        112 & 1.9653    &      829     &  0.9812    &        378    &    11.4731   &   0.7674    &                   &   1.0551    &    0.1981 \\
LMMmax  & 0.0260  &          0 & 1.9653    &       18     &  0.9819    &        397    &    10.3121   &   5.1189    &                   &   1.0551    &    0.2023 \\ \hline
        & \multicolumn{6}{c|}{mean estimated parameters and type I error} & \multicolumn{3}{c|}{estimated std. deviations} & \multicolumn{2}{c|}{variances of estimated coefficients} \\ \hline
        & $\beta$ & n($\beta$) & $\beta_w$ & n($\beta_w$) & $\beta_b$  & n($\beta_b$)  &     $\sigma$ &  $\sigma_b$ &  $\sigma_t$       &   VAR[$\beta_w$] & VAR[$\beta_b$] \\ \hline
LMMsine & 0.0132 &     8 & -0.0085 &     9 & -0.0188 &    19 &    9.9984  & 0.9696 &                          & 0.4211 & 0.1981 \\ 
GAMMfs  & 0.0215 &    13 & -0.0250 &     9 & -0.0188 &    19 &    9.9770  & 0.9722 & 5.6891                   & 0.6328 & 0.1981 \\ 
GAMMby  & 0.0215 &    13 & -0.0250 &     9 & -0.0188 &    19 &    9.9770  & 0.9722 & 5.6891                   & 0.6328 & 0.1981 \\ 
LMMmin  & 0.0263 &   112 & -0.0347 &   377 & -0.0188 &    19 &   11.4731  & 0.7674 &                          & 1.0551 & 0.1981 \\ 
LMMmax  & 0.0260 &     0 & -0.0347 &     0 & -0.0181 &    18 &   10.3121  & 5.1189 &                          & 1.0551 & 0.2023 \\ \hline
\end{tabular}
}
\end{sidewaystable}

\begin{sidewaystable}[p]
\centering
\caption{Mean estimated coefficients and number of simulation runs out of a total of 1000 in which an effect is significant at $\alpha = 0.01$, for models fit to {\bf data with sine waves with varying phases}.  In the generating model, $\beta = 0, \beta_w = 2, \beta_b = 2, \beta_{bw} = -3, \alpha = 8, \sigma_\phi = 2, \sigma = 16, \sigma_b = 1, \sigma_{b_w} = 1$.  For the simulations for the Type~I Error rate, $\beta_w = \beta_b = 0 = \beta_{bw} = 0$. } \vspace*{0.7\baselineskip}
\label{tab:phasesim1000}
{\footnotesize 
\begin{tabular}{|l|rrrrrrrr|rrrr|rrr|} \hline
& \multicolumn{8}{c|}{mean estimated parameters and power} & \multicolumn{4}{c|}{estimated standard deviations} & \multicolumn{3}{c|}{variances of estimated coefficients}  \\ \hline
         & $\beta$   & n($\beta$)   
         & $\beta_w$ & n($\beta_w$) 
         & $\beta_b$ & n($\beta_b$) 
         & $\beta_{bw}$ & n($\beta_{bw}$) 
         & $\sigma$ & $\sigma_b$ & $\sigma_{b_{w}}$ & $\sigma_t$ 
         & VAR[$\beta_w$] & VAR[$\beta_b$] &  VAR[$\beta_{bw}$] \\ \hline
  GAMMfs &    0.0157 &            7 & 2.0833 &   551     & 1.9657  &        422 &     -2.7625 &            494 &  15.9512 & 0.8650 & 0.5772 & 6.5188       & 0.5213 & 0.5653 & 1.0087 \\ 
  GAMMby &    0.0157 &            7 & 2.0833 &   551     & 1.9657  &        422 &     -2.7625 &            494 &  15.9512 & 0.8651 & 0.5773 & 6.5188       & 0.5213 & 0.5653 & 1.0087 \\ 
  LMMmax &    0.0102 &           11 & 2.0943 &   489     & 1.9733  &        427 &     -2.7776 &            442 &  16.9633 & 0.8077 & 1.1718 &              & 0.5757 & 0.5889 & 1.1135 \\ \hline
& \multicolumn{8}{c|}{mean estimated parameters and type I error} & \multicolumn{4}{c|}{estimated standard deviations} & \multicolumn{3}{c|}{variances of estimated coefficients}  \\ \hline
         & $\beta$   & n($\beta$)   
         & $\beta_w$ & n($\beta_w$) 
         & $\beta_b$ & n($\beta_b$) 
         & $\beta_{bw}$ & n($\beta_{bw}$) 
         & $\sigma$ & $\sigma_b$ & $\sigma_{b_{w}}$ & $\sigma_t$ 
         & VAR[$\beta_w$] & VAR[$\beta_b$] &  VAR[$\beta_{bw}$] \\ \hline
GAMMfs & 0.0157 &     7 & 0.0833 &     8 & -0.0343 &     4 & 0.2375 &    10 &  15.9512 & 0.8650 & 0.5772 & 6.5188 & 0.5213 & 0.5653 & 1.0087 \\ 
GAMMby & 0.0157 &     7 & 0.0833 &     8 & -0.0343 &     4 & 0.2375 &    10 &  15.9512 & 0.8651 & 0.5773 & 6.5188 & 0.5213 & 0.5653 & 1.0087 \\ 
LMMmax & 0.0102 &    11 & 0.0943 &     7 & -0.0267 &     9 & 0.2224 &     7 &  16.9633 & 0.8077 & 1.1716 &        & 0.5757 & 0.5889 & 1.1135 \\ \hline
\end{tabular}
}
\end{sidewaystable}

We fitted three models, GAMMfs, GAMMby, and LMMmax. Given that by-subject random slopes were part of the generating model, the model formulae of GAMMfs and GAMMby also include this random effect: {\tt s(Subject, Factor\_Within, bs="re")}. Simulation results of 1000 runs are presented in Table~\ref{tab:phasesim1000}.  The power of the different models is roughly the same, except for the within-subject treatment, where the power of the LMMmax is substantially reduced.  In addition, LMMmax's estimates of $\sigma$ and $\sigma_b$ are also slightly less accurate than those of the GAMMs; however, LLMmax provides a better estimate of $\sigma_{b_w}$. With respect to Type I errors, all models show nominal performance.  Finally, GAMMs emerge with lower variances for all the treatment effects, indicating that a properly specified GAMM offers researchers improved opportunities for obtaining consistent results across replication studies.

{\color{black}
\subsection{Simulations with the parameters of \citet{thul2021using}}

In the preceding simulations, we have used parameter settings that are different from those of \citet{thul2021using}, as this allows us to verify that the problem they observed generalizes.  In what follows, we report two simulations studies using the parameter settings used by \citet{thul2021using}, and using their software to generate the datasets, to illustrate that with their code and parameter settings, the problem they reported is no longer present when a more recent version of the {\bf mgcv} package is used.}

{\color{black}We used the {\tt sim\_2x2} function from the {\bf autocorr} package to generate 100 datasets for the first scenario of \citet{thul2021using}, i.e., sine with varying phase. Given that in this scenario, the power difference between LMM and GAMM is the largest with the condition of maximum effect size, for both the randomized and blocked designs (Figure 6 and 7 in \citet{thul2021using}), following their parameter settings, we set the effect size for the within-subject factor to 0.25, for the between-subject treatment to 0.5, and for the interaction to 0.5. Subsequently we fitted LMMs and GAMMs\footnote{\citet{thul2021using} considered two kinds of factor smooths, one in which the linear basis function is not penalized (directive {\tt m} is set to 2), and one in which it is (directive {\tt m} is set to 1). In this simulation, we set {\tt m} to 2.} with the {\tt fit\_2x2} function, once with an older version of {\bf mgcv} (1.8-31), and the other time with version 1.8-36. 

The upper part of Table~\ref{tab:thul_data_fit} shows that, regardless of alpha levels and for both the randomized and blocked designs, GAMMs (fitted with the old version) slightly outperform LMMs for the within-subject treatment and interaction effects, but fail more often to detect the between-subject effect, replicating the results reported by \citet{thul2021using}. When we fitted the same datasets with the new version of {\bf mgcv} (where the orthogonalization problem of factor smooths is fixed), the power of GAMMs becomes the same as that of LMMs, as shown in the lower part of Table \ref{tab:thul_data_fit}. In addition, the variance of the $\beta_b$ is also now much lower than it was, and is now comparable to the estimates of LMM.}

\begin{table}[]
\caption{Mean estimates and power (in counts out of 100 simulation runs) for a LMM and (unpenalized) GAMM, fitted to simulated datasets generated with the functions developed by \citet{thul2021using}. Parameter settings are $\beta_w = 0.25, \beta_b = 0.5, \beta_{bw} = 0.5$. $\alpha$ is set to either 0.05 or 0.01; power is presented as the count of simulations in which an effect is reported to be significant (0.05/0.01).  The upper part of the table lists results obtained with an older version of {\bf mgcv} (1.8-31), whereas the lower part presents results obtained with {\bf mgcv} version 1.8-36. The numbers in bold highlight that the low power for between-subject treatment effects is fixed with the new version.}
\label{tab:thul_data_fit}
\centering
{\footnotesize
\begin{tabular}{|cl|rrrrrr|rrr|} \hline
& & \multicolumn{6}{c|}{estimated parameters and power} & \multicolumn{3}{c|}{variances of estimated coefficients}  \\ \hline
         & & $\beta_w$ & n($\beta_w$) 
         & $\beta_b$ & n($\beta_b$) 
         & $\beta_{bw}$ & n($\beta_{bw}$) 
         & VAR[$\beta_w$] &  VAR[$\beta_b$] &  VAR[$\beta_{bw}$] \\ \hline
  \multirow{ 2}{*}{randomized} &
  LMM & 0.244  &  62/34 & 0.539 &  78/47 & 0.536 &  67/47 & 0.012 & 0.041 & 0.058 \\ 
  & GAMM & 0.248 &  66/38 & 0.518 &  {\bf 49/24} & 0.530 &  70/46 & 0.012 & 0.085 & 0.058 \\ \hline
  \multirow{ 2}{*}{blocked} &
  LMM & 0.221 &  20/14 & 0.539 &  78/47 & 0.552 &  31/13 & 0.053 & 0.041 & 0.176 \\ 
  & GAMM & 0.251  &  60/34 & 0.520 &  {\bf 50/25} & 0.534 &  61/45 & 0.015 & 0.085 & 0.061 \\ \hhline{|==|======|===|}
    \multirow{ 2}{*}{randomized} &
  LMM & 0.244  &  62/34 & 0.539 &  78/47 & 0.536 &  67/47 & 0.012 & 0.041 & 0.058 \\ 
  & GAMM & 0.248 &  67/39 & 0.539 &  {\bf 78/47} & 0.530 &  70/46 & 0.012 & 0.041 & 0.058 \\ \hline
  \multirow{ 2}{*}{blocked} &
  LMM & 0.221 &  20/14 & 0.539 &  78/47 & 0.552 &  31/13 & 0.053 & 0.041 & 0.176 \\ 
  & GAMM & 0.248  &  55/34 & 0.539 &  {\bf 78/47} & 0.535 &  64/42 & 0.015 & 0.041 & 0.063 \\ \hline
\end{tabular}
}
\end{table}

\subsection{From sine waves to random time-varying effects}

\begin{figure}[htp]
\centering
\includegraphics[width=0.8\textwidth]{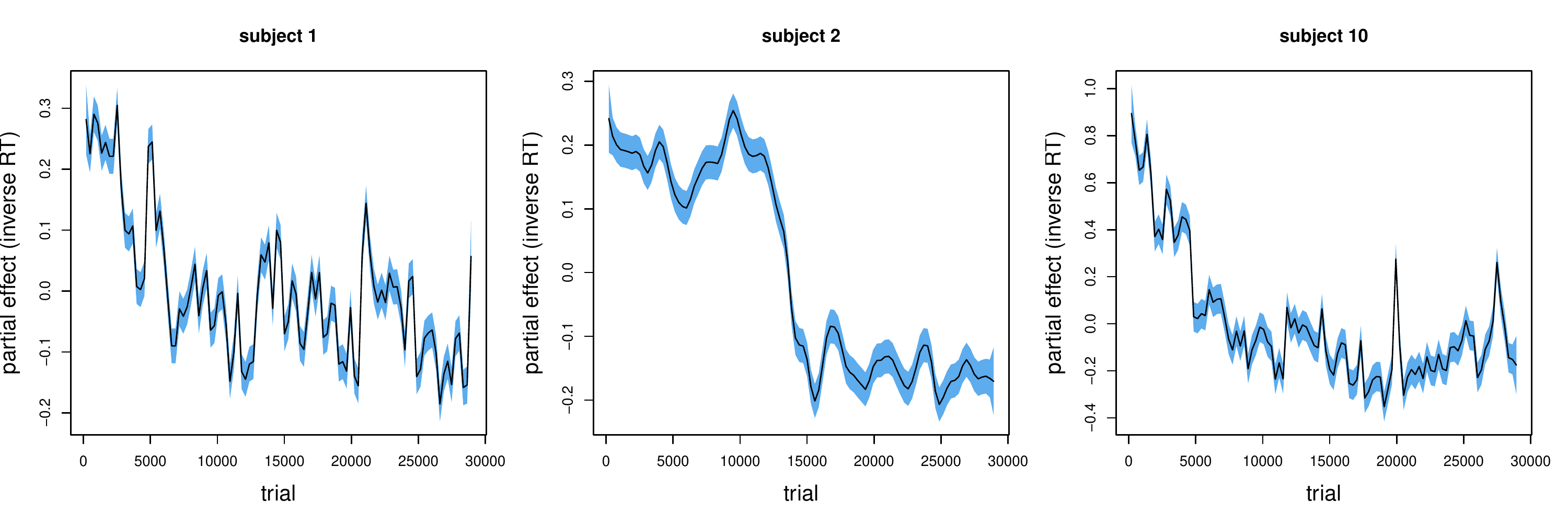}
\caption{
  Partial effects for the factor smooths fitted for three subjects in the British Lexicon Project. Note that the scale of the Y-axis is much wider for subject 10 (right panel).
}
\label{fig:time_curves}
\end{figure}

In the preceding discussion, we adopted the characterization of time-varying effects in psychometric data {\color{black} implemented in some of the key simulations reported in \citet{thul2021using}, the sine wave.  However, we have never encountered regular sinusoid trends in experimental time for real data.}  Actual time-varying effects look very different from sine waves, as illustrated in Figure~\ref{fig:time_curves} for the data of three subjects in the British Lexicon Project. None of these temporal effects is well-charactized by a sine wave.  Furthermore, subjects~1 and~2 (left and center panels) have quite small partial effects compared to subject~10 (right panel, note the difference in the scales on the Y-axis).  In addition, there is more jitter in the curve for subject~1 as compared to the curve of subject~2. All three subjects tune into the task during the first 10,000 trials. Subjects~1 and~10 become faster in a gradual way, whereas subject~2 speeds up performance rapidly in a relatively narrow time interval of some 2,500 trials.

\begin{figure}[t]
\centering
\caption{Empirical and simulated wiggly curves.  Left: Partial effects for the factor smooths for the 48 subjects in a primed auditory lexical decision experiment on Taiwan Mandarin with 380 trials.  Right: Random curves generated with the standard deviation for the weights on the basis functions set to 2.}
\label{fig:wigglies}
\includegraphics[width=0.48\textwidth]{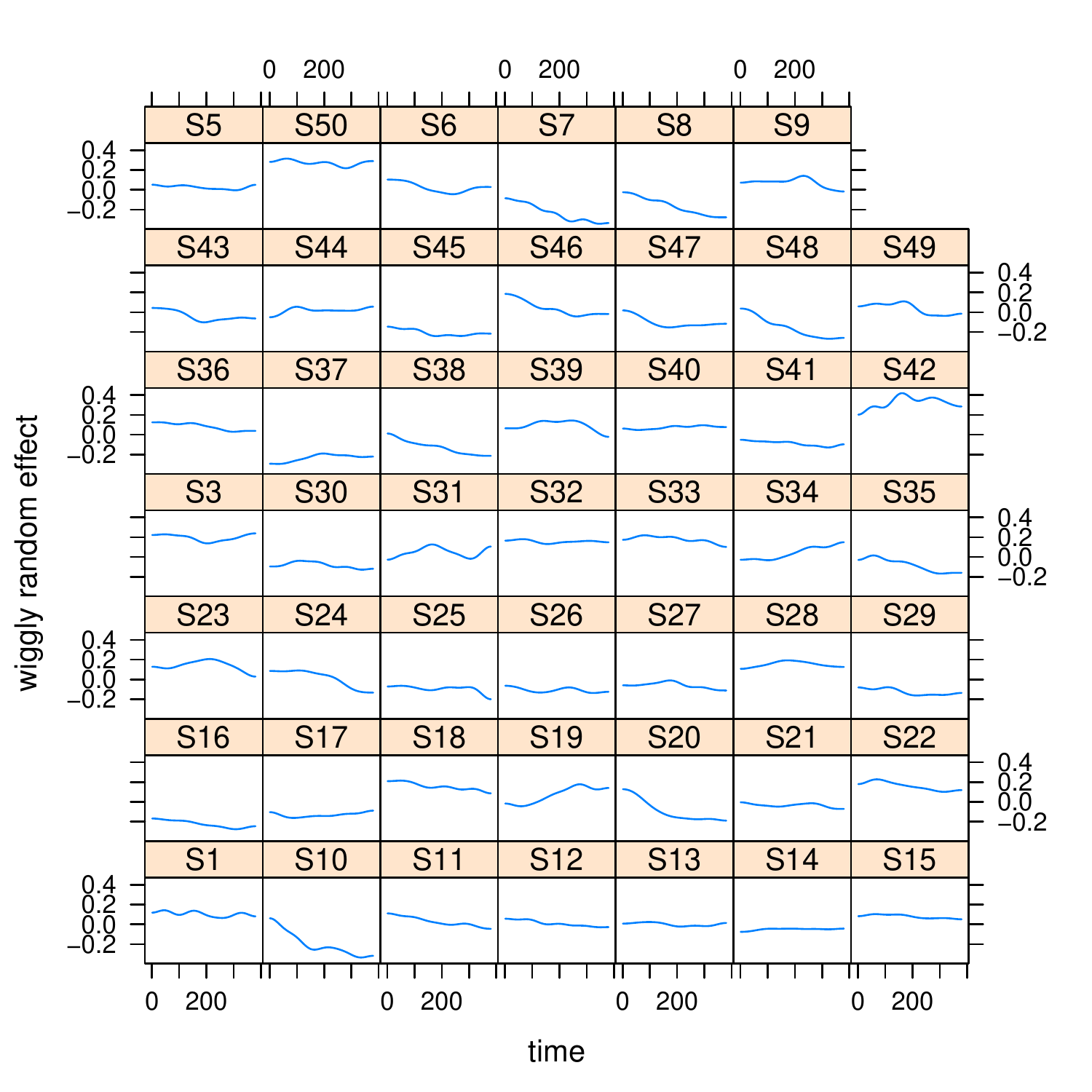}\includegraphics[width=0.48\textwidth]{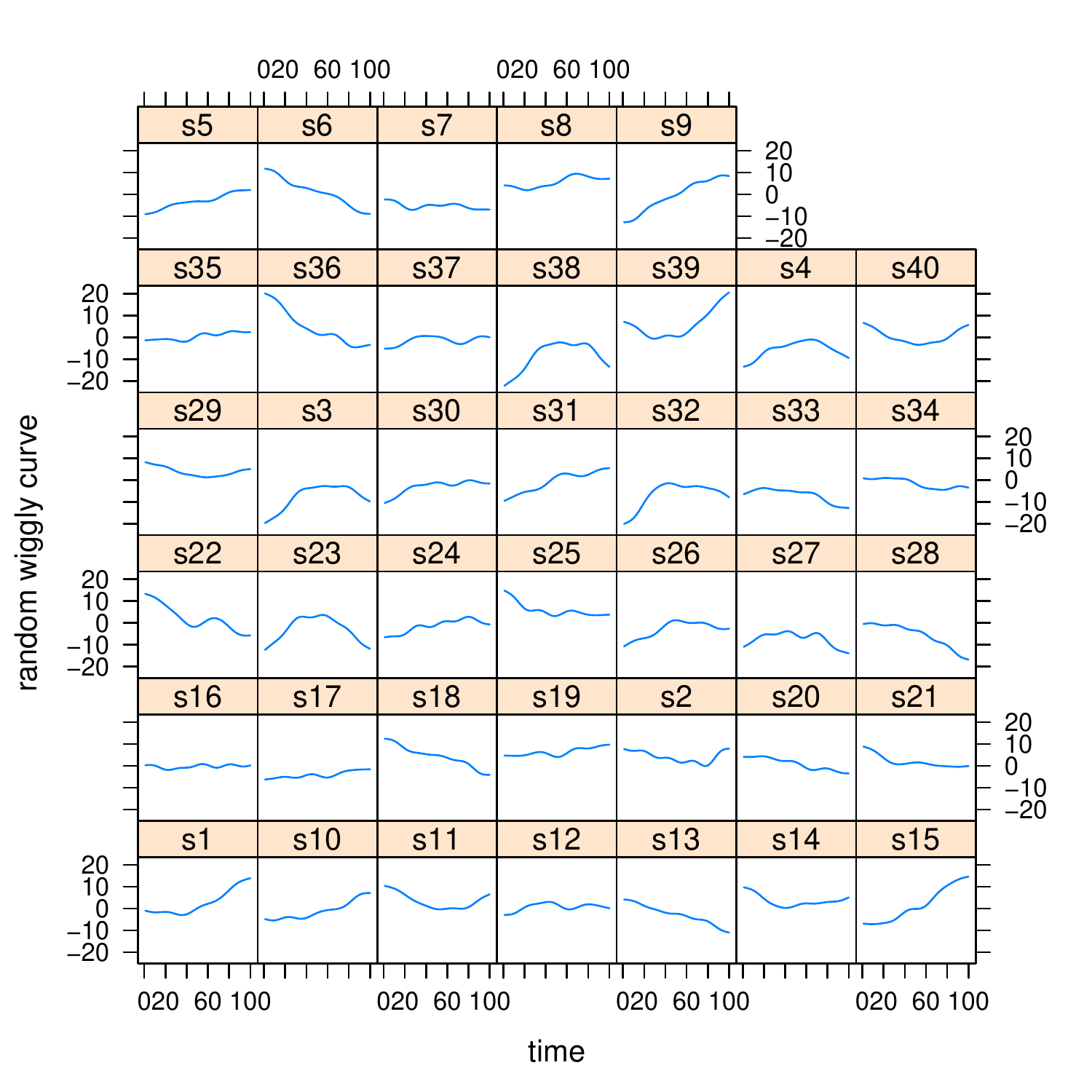}
\end{figure}

At shorter time scales, time-dependent effects can also be highly variable, as illustrated in the left panel of Figure~\ref{fig:wigglies}.  This trellis graph presents the partial effects of factor smooths for 48 subjects in a primed auditory lexical decision experiment with 380 trials \citep[see][for further details]{Chuang:2017,Chuang:Fon:Papakyritsis:Baayen:2020}.  The vertical positioning of the curves provides an indication of whether subjects are fast or slow responders.  Some subjects are quite stable over time, others show undulating patterns, some become slower near the end of the experiment, whereas yet others speed up as the experiment progresses.

It is straightforward to simulate datasets with random time-varying trends by first constructing $k$ basis functions, then weighting these by (possibly scaled) coefficients sampled from a normal distribution with mean zero, and finally adding them to obtain a random wiggly curve (the supplementary materials include R code).  The right panel of Figure~\ref{fig:wigglies} illustrates the kind of curves generated by this function.

\newpage

\begin{sidewaystable}[p]
\centering
\caption{Mean estimates, power and type I error rate (in counts out of 1000 simulation runs) for a GAMM with a factor smooth for time (GAMMfs), a GAMM using by-smooths (GAMMby), and a maximal linear mixed model (LMMmax), fitted to {\bf simulated datasets with random wiggly curves}.  The parameters of the model generating the data are $\beta=0, \beta_w = 1, \beta_b = 2, \beta_{bw} = -3,  \sigma=10, \sigma_{b}=4, \sigma_{b_w} = 2$.  The standard deviation for the random weights for the basis functions $\sigma_{\text{tprs}}$ was set to 2. 
} \vspace*{0.6\baselineskip}
\label{tab:wiggly}
{\footnotesize
\begin{tabular}{|l|rrrrrrrr|rrrr|rrr|} \hline
& \multicolumn{8}{c|}{mean estimated parameters and power} & \multicolumn{4}{c|}{estimated standard deviations} & \multicolumn{3}{c|}{variances of estimated coefficients}  \\ \hline
         & $\beta$   & n($\beta$)   
         & $\beta_w$ & n($\beta_w$) 
         & $\beta_b$ & n($\beta_b$) 
         & $\beta_{bw}$ & n($\beta_{bw}$) 
         & $\sigma$ & $\sigma_b$ & $\sigma_{b_{w}}$ & $\sigma_t$ 
         & VAR[$\beta_w$] &  VAR[$\beta_b$] &  VAR[$\beta_{bw}$] \\ \hline
  GAMMfs & -0.0243 &     8 & 1.2009 &   164 & 1.9987 &   115 & -2.7872 &   772 &  9.9619  & 3.9342 & 1.3850 & 1.1670 & 0.1954 & 1.7667 & 0.3950 \\ 
  GAMMby & -0.0243 &     8 & 1.2009 &   164 & 1.9987 &   115 & -2.7872 &   772 &  9.9619  & 3.9342 & 1.3850 & 1.1670 & 0.1954 & 1.7667 & 0.3950 \\ 
  LMMmax & -0.0273 &    14 & 1.2070 &   155 & 2.0040 &   159 & -2.7979 &   694 &  11.1639 & 3.8977 & 1.9583 &        & 0.2377 & 1.8032 & 0.5005 \\ \hline
& \multicolumn{8}{c|}{mean estimated parameters and type I error} & \multicolumn{4}{c|}{estimated standard deviations} & \multicolumn{3}{c|}{variances of estimated coefficients}  \\ \hline
         & $\beta$   & n($\beta$)   
         & $\beta_w$ & n($\beta_w$) 
         & $\beta_b$ & n($\beta_b$) 
         & $\beta_{bw}$ & n($\beta_{bw}$) 
         & $\sigma$ & $\sigma_b$ & $\sigma_{b_{w}}$ & $\sigma_t$ 
         & VAR[$\beta_w$] &   VAR[$\beta_b$] &  VAR[$\beta_{bw}$] \\ \hline
  GAMMfs & -0.0243 &     8 & -0.7991 &    40 & -0.0013 &     9 & 0.2128 &     1 &  9.9619 & 3.9342 & 1.3850  & 1.1670   &  0.1954 & 1.7667 & 0.3950 \\ 
  GAMMby & -0.0243 &     8 & -0.7991 &    40 & -0.0013 &     9 & 0.2128 &     1 &  9.9619 & 3.9342 & 1.3850  & 1.1670   &  0.1954 & 1.7667 & 0.3950 \\ 
  LMMmax & -0.0273 &    14 & -0.7930 &    38 &  0.0040 &    15 & 0.2021 &     1 &  11.1639 & 3.8977 & 1.9583 &          &  0.2377 & 1.8032 & 0.5005 \\ \hline
\end{tabular}
}
\end{sidewaystable}

\clearpage

To clarify potential advantages or disadvantages of using GAMMs or LMMs for data with wiggly random effects, we simulated datasets in which the within-subject treatment was not blocked, as this is the design for which \citet{thul2021using} recommend using the LMM rather than the GAMM.  The parameters of the model generating the data are $\beta = 0, \beta_w = 1, \beta_b = 2, \beta_{bw} = -3,  \sigma=10, \sigma_{b}=4, \sigma_{b_w} = 2$.  The standard deviation for the random weights for the basis functions $\sigma_{\text{tprs}}$ was set to 2.  
Table~\ref{tab:wiggly} reports performance, evaluated over 1000 simulation runs,  of three models,  a GAMM with factor smooths (GAMMfs), a GAMM with by-smooths (GAMMby), and a maximal linear mixed model (LMMmax).  

Type I error rates are nominal, except for the within-subject treatment: although fine at $\alpha=0.05$, they are too high for $\alpha=0.01$ for all three models.  Power for the within-subject treatment is slightly superior for the GAMMs.  With respect to the between-subject treatment, here the power of GAMMs lags behind that of LMMmax, in line with what \citet{thul2021using} observed for their simulations.   However, for the interaction of the two treatments, the reverse holds: here, the GAMMs outperform LMMmax, without compromising on the type~I error rate.     The right-hand side of the table shows that again GAMMs  offer estimated effect sizes that are less variable across simulations.   With respect to the estimates of the standard deviation parameters, the GAMMs are more precise for $\sigma$ and $\sigma_b$, but LMMmax does better for $\sigma_{b_w}$.  Overall, GAMMs emerges as a good choice, a choice that comes with the advantage of offering the analyst insight into the time-varying trends in her data.

\section{Interactions with time in experimental data}

{\color{black}Thus far, we have considered simulated data only. \citet{thul2021using} (in contrast to \citet{BarrGam:2020}) also report some simulations modeled on empirical experimental data. In what follows, rather than adding more simulation studies seeking to approximate actual datasets, we illustrate what can be achieved when GAMMs are applied to actual experimental data, moving from analysis of variance to multiple regression, as it is here that GAMMs truly come into their own.}

How important is it to have insight into time-varying trends in chronometric data? According to \citet{thul2021using},  time-varying effects, which they dub ``nuisance variation'', can occasionally be of interest in their own right, but more often they can be treated as irrelevant and ignored.  They argued that GAMMs come into their own when the experimental trials themselves are time series, as is the case for, e.g., pupil dilation curves  \citep{van2019analyzing}.  In this section, we argue that also for simple sequences of response variables, investigating time-varying effects can yield valuable insights. 

\begin{figure}[ht]
\centering
\includegraphics[width=0.8\textwidth]{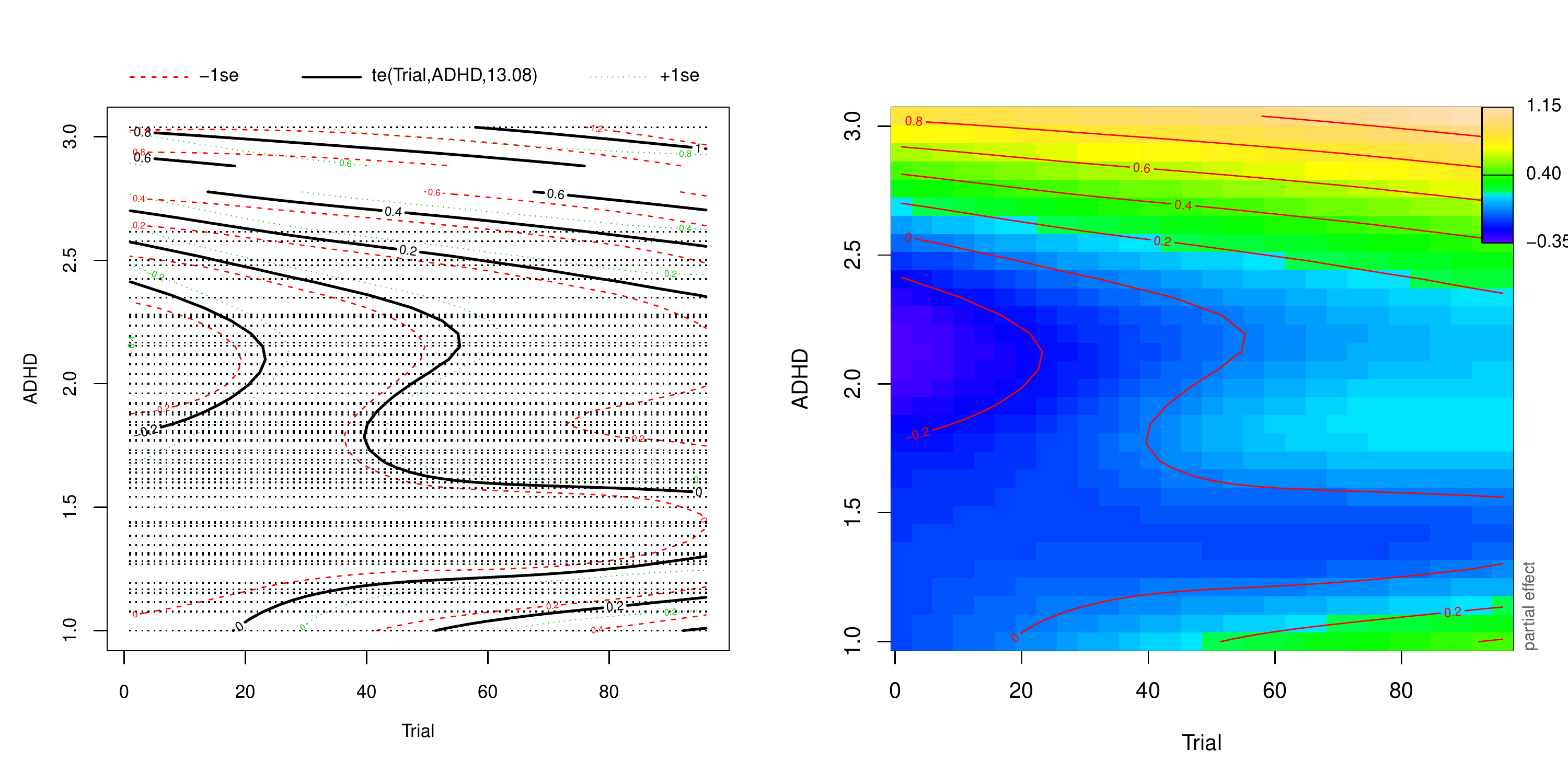}
  \caption{Interaction of Trial by ADHD-score in a model fitted (using a tensor product smooth) to the minimum curvature of swipe movements in an arithmetic task. In the left panel, the colored lines indicate 1SE confidence regions, in the right panel, darker colors denote lower values of minimum curvature, and brighter, more yellow colors indicate higher values of the response.}
  \label{fig:swipe}
\end{figure}

\citet{Baayen:Vasishth:Kliegl:Bates:2017} observed for one dataset that subjects with more wiggly curves made more errors.  Variability in performance can have many sources, ranging from the level of skilled performance and practice \citep{segalowitz1993skilled} to working memory capacity and degree of ADHD.  \citet{mock2018predicting} observed increasing variability in performance for subjects with higher values on a scale measuring ADHD.  They conducted an experiment requiring subjects to respond to mathematical problems using a touchpad.  One measure they considered was the minimal curvatures of the swipes made.  Minimal curvatures increased in the course of the experiment.  Interestingly, these changes were greater for subjects with higher ADHD scores, {\color{black} as can be seen in contour plots presented in Figure~\ref{fig:swipe}. Contour lines connect data points with the same expected value of the response variable. The left panel adds 1SE confidence regions around these contour lines, and the right panel presents the fitted values using color coding, in the same way as familiar from topographic maps. The effect of trial for a given value of ADHD can be assessed by tracing a horizontal line at the desired ADHD value. For ADHD=1.5, no contour lines are crossed, hence there is little effect of trial for subjects with this score.  Conversely, for ADHD=2.75, several contour lines are crossed, with a positive gradient, indicating that subjects with this ADHD score evidenced increasingly large values of the response variable as the experiment proceeded.} In other words, GAMMs may be of use to researchers interested in individual differences and how such differences lead to different response patterns over experimental time.

\begin{figure}[htbp]
\centering
\includegraphics[width=1.0\textwidth]{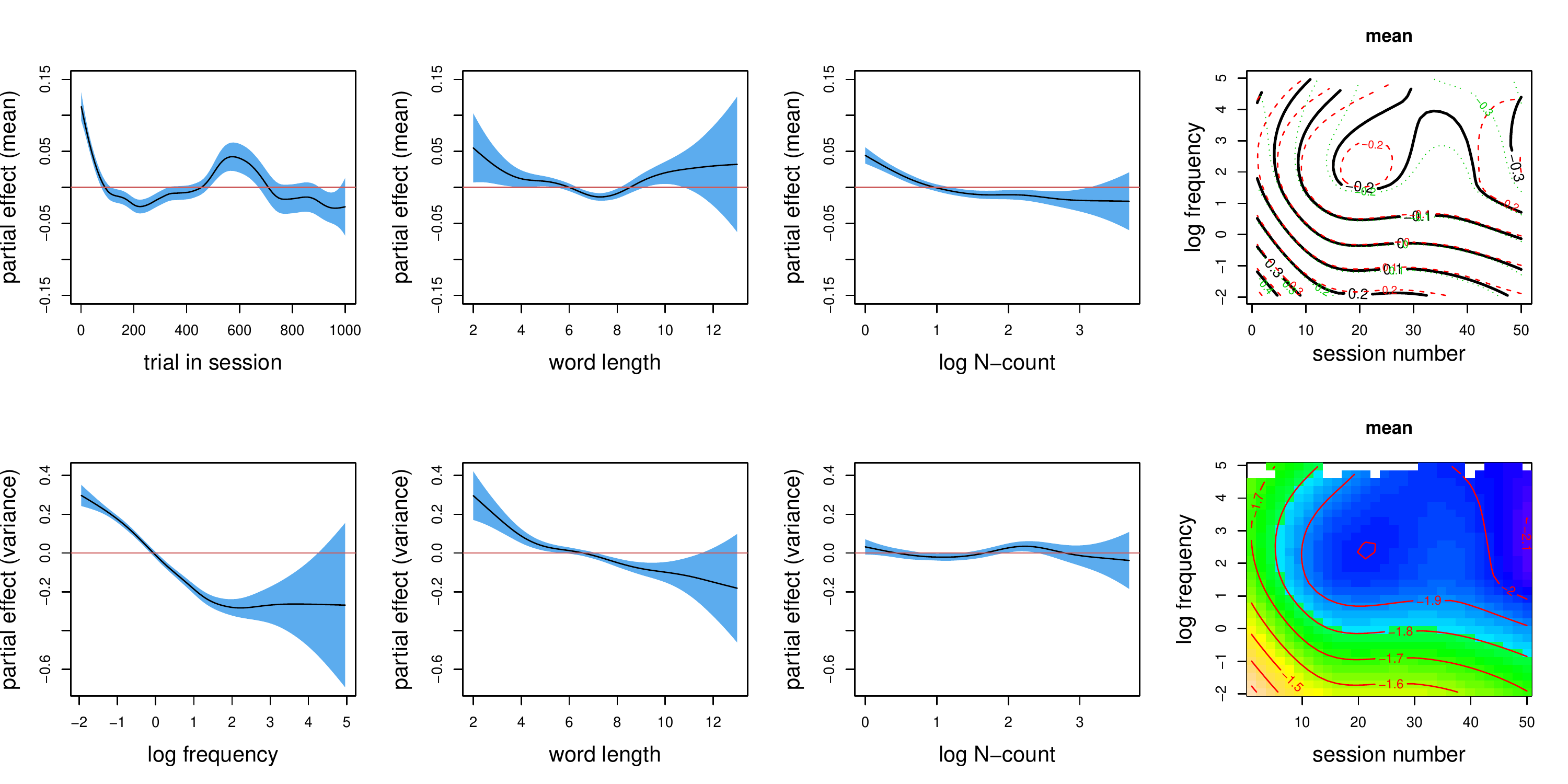}
\caption{Partial effects in a Gaussian Location Scale GAM fitted to the inverse-transformed reaction times to words of subject 1 in the British Lexicon Project.  In the contour plots, darker colors indicate shorter reaction times. In the upper right, green contour lines specify 1 SE up, and red lines 1 SE down from their respective contour lines in black.}
\label{fig:blp1}
\end{figure}

As an example of a dataset with reaction time as response variable that illustrates this point, consider again the data of subject 1 in the British Lexicon Project.  Reaction times were collected during 50 sessions, so we included a smooth for session, and trial number within session as temporal predictors.  A Gaussian Location-Scale GAM was fitted to the inverse-transformed responses to word trials, with as lexical predictors frequency of occurrence ({\tt log frequency}), length in letters ({\tt word length}), and neighborhood size ({\tt log N-count}).  Figure~\ref{fig:blp1} presents the partial effects of the predictors, and Table~\ref{tab:blp1} provides the model summary.  The effects of word length and neighborhood density on mean reaction time are small compared to the large effect on the mean of frequency, which interacted with session number.  This interaction is visualized in the rightmost panels of Figure~\ref{fig:blp1}.  Reaction times decreased over the course of the sessions.  In the early sessions, there was a strong U-shaped effect of frequency; {\color{black} for instance, for session 10, when moving parallel along the vertical axis, a negative gradient is followed by a positive gradient. Thus, although the lowest-frequency words elicited the longest reaction times in this session,  somewhat longer reaction times are present for very high-frequency words as compared to intermediate frequency words.} In later sessions, the U-shaped effect disappeared, leaving a smaller-sized frequency effect that leveled off for words with above-average frequency.   The variance of reaction times {\color{black}(visualized in the bottom panels of Figure~\ref{fig:blp1}}),  decreased substantially with frequency, and to a lesser extent with length.  Variance in RTs was not predictable from a further interaction of frequency by session number. 

The interaction of session number by frequency is of theoretical interest, as it challenges theories that model the word frequency effect by means of fixed resting activation levels of word units.  Such theories will need to consider response biases that change over the many sessions of mega-experiments, as participants are seeking to optimize their word/non-word decision making.  In connectionist approaches \citep[see, e.g.,][]{Harm:Seidenberg:2004}, changes in the effect of frequency might be argued to be a consequence of continued learning.  

\begin{table}[ht]
\centering
\caption{Model summary of a Gaussian Location-Scale GAMM fitted to the inverse-transformed reaction times of subject 1 to words in the British Lexicon Project.} 
\begin{tabular}{lrrrr} \hline
A. parametric coefficients & Estimate & Std. Error & t-value & p-value \\ \hline
  Intercept (mean)      & -1.7367 & 0.0019 & -932.6938 & $<$ 0.0001 \\ 
  Intercept (variance)  & -1.5717 & 0.0062 & -253.7814 & $<$ 0.0001 \\ \hline
B. smooth terms (mean) & edf & Ref.df & F-value & p-value \\ \hline
  s(trial within session) & 11.4179 & 14.2413 & 251.7092 & $<$ 0.0001 \\ 
  te(session, log frequency) & 18.3005 & 20.5734 & 5281.5619 & $<$ 0.0001 \\ 
  s(word length) & 4.7944 & 5.8089 & 31.3746 & $<$ 0.0001 \\ 
  s(log N-count) & 3.5946 & 4.4088 & 65.4796 & $<$ 0.0001 \\ \hline
C. smooth terms (variance) & edf & Ref.df & F-value & p-value \\ \hline  
  s(log frequency) & 4.3186 & 5.2081 & 632.8609 & $<$ 0.0001 \\ 
  s(word length) & 3.9879 & 4.9434 & 49.5434 & $<$ 0.0001 \\ 
  s(log N-count) & 4.0608 & 4.9153 & 12.4410 & 0.0300 \\ \hline
\end{tabular}
\label{tab:blp1}
\end{table}

\section{Discussion}

\citet{BarrGam:2020} called attention to problems that arise when modeling time-varying effects in experiments with between-subject and within-subject treatment factors using generalized additive models with factor smooths.   We have shown that these problems do not arise when a `by'-smooth is used, and that from {\bf mgcv} version 1.8-36 onwards, these problems also no longer occur with GAMMs using factors smooths.  

\citet{BarrGam:2020} claimed that the linear mixed model is safe to use for counterbalanced data, even in the presence of time-varying effects.   We have presented a series of simulation studies showing that once the smooths for time-varying effects are properly specified,  GAMMs show excellent performance, on a par with or better than the LMM.   GAMMs also perform very well for blocked designs,  where the LMM suffers catastrophic loss of power. {\color{black} \citet{thul2021using} {\color{black} observed that when trial-to-trial dependencies exist, blocked designs can be problematic for LMMs, and that researchers therefore should make sure to proper counterbalance}. However, for some research questions, blocked designs may actually be necessary.  GAMMs thus enable running experimental designs that cannot be properly analysed with the LMM, if strong time-varying effects are present in the data.} In addition, GAMMs {\color{black} may} offer estimates of treatment effects that are less variable across simulation runs than those produced by LMMs.  This makes GAMMs especially attractive for replication studies.  From a methodological perspective, this finding highlights the danger inherent in judging a statistical model not by its faithfulness to the forces shaping the data, but by power and type~I error: both power and error rate may look good, {\color{black} but when the true complexity of the data is left unexplored, as illustrated by the present example simulations, results across replication experiments run the risk of being less consistent than they need be}.

\citet{thul2021using} argued that time-varying effects in chronometric experiments {\color{black} ``can usually be safely ignored" (p. 15) when time-varying effects are not of interest.}  However,  recommending a modeling strategy that makes it impossible for the field to detect time-varying effects seems ill-advised.  We have therefore presented an example in which a strong predictor for reaction times in the visual lexical decision task, word frequency, interacted with experimental time in a theoretically informative way. 

Finally, if a mathematical function is available that describes a non-linear effect, the non-linear mixed model \citep[cf.][]{Pinheiro:Bates:2000}  may provide more precise fits, although the algorithms of the {\bf nlme} package are stretched beyond their limits for the sinusoid time-varying effects present in our simulations.  We are indebted to Douglas Bates for pointing out to us that the varying-amplitude sine waves in the simulations of \citet{thul2021using} are straightforward to model using LMMs.  Our simulations show that prior knowledge of the functional form of the time-varying effect leads to improved results compared to GAMMs.  However,  since mathematical models predicting the complex ways in which subjects go through an experiment over time are not available,  in the foreseeable future, the generalized additive (mixed) model will remain a useful tool for coming to grips with {\color{black} time-varying effects in  experimental data}. 

{\color{black}
\citet{thul2021using} raise a more general issue, namely, that GAMs would come with too many researcher degrees of freedom, that the costs of their use is therefore not well justified, and that hence the field is better served by restricting itself to the LMM, at least for the analysis of behavioral experiments with factorial predictors and between-subject designs.  It is understandable that a field that is plagued by a replicability crisis \citep{francis2012publication,open2015estimating} is eager to embrace fixed procedures for the statistical evaluation of its data.  Nevertheless, one-size-fits-all procedures come with their own disadvantage.  Limiting statistical analysis to the linear mixed model blocks progress, even for simple datasets with factorial predictors.  Researchers will be discouraged to test for potential time-varying effects and possible interactions with predictors of interest, and as a consequence, empirical evidence on whether such effects exist, and if they exist, what their prevalence is, will not be forthcoming.\footnote{For clarity, we note that for datasets where issues of temporal dependencies, and potentially non-linear covariates, are not at issue, the linear mixed model is an excellent choice, and we recommend the MixedModels package for Julia as offering the most optimized implementation currently available.}

We note here that lexical predictors that are numeric rather than categorical often have non-linear effects, and that, as demonstrated in the present study, such nonlinear effects may interact with experimental time in theoretically informative ways.  \citet{thul2021using} may well be right that interactions of experimental time with factorial variables may be rare, but they cannot be ruled out in principle, and with the updated code, it is straightforward to clarify for a given dataset whether time-varying effects are present, and whether they interact or not.  

For a field to make innovative scientific progress, one of the things it has to do is encourage its researchers to explore experimental data with novel statistical methods that enable improved insight into the full complexities of experimental datasets.  Novel findings about non-linearities will then naturally be followed by further studies seeking to replicate these non-linearities and the details of their functional form.   
        
But what if GAMMs come with `excessive researcher degrees of freedom',  and that therefore replication studies with GAMMs are not viable?  In response to this concern of \citet{thul2021using}, we first note that in practice, the choices available to the analyst when using GAMMs are highly constrained by the data under investigation.  Indeed, GAMMs have been developed in response to analytical needs across a wide range of sciences; psychology and linguistics are newcomers and, as usual, late adapters.  Given the interdisciplinary needs driving the development of GAMMs, unsurprisingly, GAMMs offer the analyst a wide spectrum of tools for understanding data. But many of these tools are not relevant for any specific dataset. For instance, smoothing on a sphere, or smoothing with isotropic thin plate regression splines, should not be used for modeling the interaction of time by frequency of use.  Here, tensor product smooths are appropriate.  Any honest and conscientious user of GAMMs will find her modeling options to be highly constrained.  
        
Second, experiences with GAMMs have clarified that starting with maximal GAMMs is ill-advised: a five-way tensor product for five covariates is likely to be completely uninterpretable.  Progress is best served by studying interactions in low dimensions; here, exploratory visualization in combination with theoretical guidance are indispensable. Once non-linearities have been detected in exploratory studies, replication studies are essential both for consolidating results, and for establishing guidelines for best practice in data analysis.  

At this point, the question arises of what may be expected to replicate. It is useful to distinguish here between two kinds of non-linearities that can be observed in psycholinguistic data.  
        
On the one hand, we may observe non-linearities that we can consider as a source of noise, such as random wiggly curves for temporal effects that play out in the course of an experiment.  To the extent that such wiggly curves are the consequence of fluctuations in attention, they are way beyond the capacity of any existing theory of attention as long as subjects are selected randomly and extensive background data on these subjects is not available.  We do not expect to be able to replicate such curves even for replication experiments with the same materials and the same subjects, as attention is likely to shift and vary at different moments in time as the experiment unfolds.  However, if the model is appropriate for the data, we may expect to find similar non-linear noise components across  replication experiments.  
      
On the other hand, we may observe non-linearities that are systematic and within reach of theoretical explanation. For instance, the effect of frequency of occurrence in lexical decision tasks has repeatedly been observed to be nonlinear.  GAMMs enable us to obtain precise predictions of the non-linear effect of frequency on reaction time, both in terms of mean and in terms of variance, and as a next step, the challenge for theories of lexical processing is to provide a mathematical explanation for why the observed specific non-linearities arise \citep[for an example of a mathematical model for a GAM regression surface, see][]{Baayen:2010}. In other words, observed systematic non-linearities that are part of the `ground truth' (and within theoretical reach) are expected to be replicable.
}

In summary, \citet{thul2021using} are correct to call attention to problems that arise when factor smooths are used that do not orthogonalize by-subject random intercepts with respect to the by-subject smooth part.  This problem has been fixed in {\bf mgcv} version 1.8-36: factor smooths now yield the same result as the corresponding `by' smooths.  However, the claim of \citet{thul2021using} that GAMMs would be ``complex, advanced techniques that are not fully understood'' and that can have ``potential side-effects'' (p. 14) is not convincing, as it is based on experience with one particular software ({\bf mgcv}) and one particular feature of that software (factor smooths, but not `by' smooths). Generalising the results of their study to all possible simulation settings,  models and software implementations of GAMMs seems quite a leap of faith.


\section{Open Practices Statement}
Data and code for this study are available at \url{https://osf.io/fbndc/}.

\bibliography{factorsmooths}

\newpage
\begin{flushleft}

{\bf Mathematical notation} \\ \vspace*{0.8\baselineskip}
\begin{tabular}{l|l}
\hline
    $\beta$        &  general intercept                                  \\
    $b$            &  by-subject random intercept                        \\
    $\beta_w$      &  within-subject effect                              \\
    $\beta_b$      &  between-subject effect                             \\
    $\beta_{bw}$   &  interaction effect between $F_b$ and $F_w$         \\
    $\alpha$       &  amplitude of sine wave                             \\
    $\phi$         &  phase shift for sine wave                          \\ 
    $\varepsilon$  &  error                                              \\ \hline
    $\sigma_b$     &  standard deviation for by-subject random intercept  \\
    $\sigma_{b_w}$ &  standard deviation for by-subject random slope for $\beta_w$     \\
    $\sigma_t^2$ & prior variance of the coefficients of the by-subject smooths, \\&  $\sigma^2/\lambda$ in $\bm{\beta} \sim {\cal N}(0, \mbox{S}^{-} \sigma^2/\lambda)$, i.e., the multiplier of the penalty \\
    $\sigma_\alpha$&  standard deviation for sine wave's amplitude        \\
    $\sigma_\phi$  &  standard deviation for sine wave's phase shift      \\
    $\sigma$       &  standard deviation for error                        \\
    $\sigma_{\text{tprs}}$&  standard deviation for the basis functions' random weights \\
\hline    
\end{tabular}

\end{flushleft}

\vspace*{2\baselineskip}
\noindent
{\bf Appendix: Factor smooth interactions and orthogonality} \\ \vspace*{0.4\baselineskip}

\noindent
Why should it be problematic to not orthogonalize the bases for penalized and unpenalized components of a factor smooth interaction term? It seems likely that this relates to the implicit assumption that the random coefficients associated with these two components are independent a \emph{priori}. To see the problem, consider the simple model
\begin{equation} \label{eq:simpModel}
y_{i} = a_{j(i)} + b_{j(i)}(x_i - \bar{x}) + \epsilon_i,
\end{equation}
where $\bar{x}$ is the mean of the $x_i$'s and $j(i)$ indicates the factor level $j \in \{1, \dots, L\}$ to which the $i$-th observation belongs. The $a_j$'s and $b_j$'s are independent, with variance $\sigma^2_a$ and $\sigma^2_b$ and zero means.  In other words, $a_{j(i)}$ denotes random intercepts, and $b_{j(i)}$ denotes random slopes. This model seems reasonable --- independence just says that we do not expect the overall mean of the $y_i$'s to be related to its rate of change with respect to $x_i$. And the orthogonality of the intercept (which is a vector of ones ${\bf 1} = [1, 1, \dots, 1]^T$) and ${\bf x} - \bar{x}$ guarantees the independence between $a_j$ and $b_j$ will be maintained in the posterior.

\noindent
But, suppose that we re-parameterize to arrive at
$$
y_{i} = a_{j(i)}' + b_{j(i)}'x_i + \epsilon_i,
$$
so that
\[
\begin{bmatrix}
    a_{j(i)}'  \\
    b_{j(i)}'
\end{bmatrix}
=
\begin{bmatrix}
    1 & -\bar{x} \\
    0 & 1 
\end{bmatrix}
\begin{bmatrix}
    a_{j(i)}  \\
    b_{j(i)}
\end{bmatrix}.
\]
The covariance matrix of ($a_j'$, $b_j'$) is therefore
\[
{\bf V}_0
=
\begin{bmatrix}
    \sigma_a^2 + \bar{x}^2\sigma_b^2 & -\bar{x}\sigma_b^2 \\
    -\bar{x}\sigma_b^2 & \sigma_b^2 
\end{bmatrix}.
\]
(For the diagonal elements, we use the equality $\mbox{Var}(aX+bY) = a^2\mbox{Var}(X) + b^2\mbox{Var}(Y)$, and for the covariances, we note that $\mbox{Cov}[a-\bar{x}b, b] = \mbox{E}[(a-\bar{x}b)-\mbox{E}[a-\bar{x}b])(b-\mbox{E}[b])] = \mbox{E}[(-\bar{x}b)(b)] = -\bar{x}\sigma_b^2$.)  Hence, while modelling $(a, b)$ as independent was reasonable, this will not be appropriate for $(a_j', b_j')$ (unless $\bar{x} = 0$). Indeed, if we decide to model $(a_j', b_j')$ as independent the covariance matrix with diagonal elements $\tilde{\sigma}^2_{a'}$ and $\tilde{\sigma}^2_{b'}$ that best matches ${\bf V}_0$ will be such that $\tilde{\sigma}_{a'}^2$ and $\tilde{\sigma}_{b'}^2$ are biased estimates of the corresponding diagonal entries of ${\bf V}_0$ (which is the true covariance matrix, that is the one used to generate the data).

Here is a practical illustration. We first simulate some data from model (\ref{eq:simpModel}):
\begin{verbatim}
set.seed(98)
n <- 100000
L <- 10000
x <- runif(n) - 0.5                    # this is x-mean(x) in (3)
g <- rep(1:L,n/L)
a <- rnorm(L)                          # sigma_a = 1
b <- rnorm(L)*0.1                      # sigma_b = 0.1
y <- a[g] + b[g] * x + rnorm(n)*.1
g <- factor(g)
\end{verbatim}
We then fit the correct model with independent random effects:
\begin{verbatim}
library(lme4)
lmer(y~(1|g)+(x-1|g))
\end{verbatim}
which outputs:
\begin{verbatim}
Linear mixed model fit by REML ['lmerMod']
Formula: y ~ (1 | g) + (x - 1 | g)
REML criterion at convergence: -101098.3
Random effects:
 Groups   Name        Std.Dev.
 g        (Intercept) 1.00258 
 g.1      x           0.09925 
 Residual             0.10010 
Number of obs: 100000, groups:  g, 10000
Fixed Effects:
(Intercept)  
    0.01275  
\end{verbatim}

The standard deviations of the random effects are close to their true values, $\sigma_a = 1$ and $\sigma_b = 0.1$. Now we shift $x$ to remove the independence between intercept and slope, but fit the same model which assumes independence between the random effects: 
\begin{verbatim}
x1 <- x+4                    # this is x in (3), with mean(x) = 4
lmer(y~(1|g)+(x1-1|g))   
\end{verbatim}
\begin{verbatim}
Linear mixed model fit by REML ['lmerMod']
Formula: y ~ (1 | g) + (x1 - 1 | g)
REML criterion at convergence: -101509.3
Random effects:
 Groups   Name        Std.Dev.
 g        (Intercept) 0.98451 
 g.1      x1          0.08087 
 Residual             0.10104 
Number of obs: 100000, groups:  g, 10000
Fixed Effects:
(Intercept)  
    0.01002      
\end{verbatim}
The variance estimates are biased downward, in fact the true random effects standard deviations are $({\bf V}_0)_{11}^{1/2} = \sqrt{\sigma_a^2 + \bar{x}^2\sigma_b^2} \approx 1.077$ and $({\bf V}_0)_{22}^{1/2} = 0.1$. The downward bias is due to the fact that the negative correlation between the random effects is being ignored. Fitting a model with correlated random effects avoids the problem:
\begin{verbatim}
lmer(y~(x1|g)) 

Linear mixed model fit by REML ['lmerMod']
Formula: y ~ (x1 | g)
REML criterion at convergence: -102017.6
Random effects:
 Groups   Name        Std.Dev. Corr 
 g        (Intercept) 1.07779       
          x1          0.09933  -0.37
 Residual             0.10009       
Number of obs: 100000, groups:  g, 10000
Fixed Effects:
(Intercept)  
    0.01265  
\end{verbatim}

How is the problem described here relevant to factor smooth and by-smooths effect fitted in ${\bf mgcv}$? In ${\bf mgcv}$ the random effect on the intercept is modelled as independent from the smooth part (which we can think of in terms of polynomials bases $x$, $x^2$, $x^3$, ...). That is, the correlation between the random intercept and the smooth part is assumed to be zero. But, if the intercept is not orthogonal to the spline basis functions used to form the smooth part, the problem described above might occur. Factor smooth interactions constructed under ${\bf mgcv}$ version lower than 1.8-36 where affected by this problem, while orthogonalization between the intercept and the rest of the smooth has always been enforced in by-smooths.

\end{document}